\documentclass[twocolumn]{aastex631}
\usepackage{natbib}
\usepackage{latexsym}
\usepackage{graphicx}
\usepackage{epsfig}
\usepackage{amssymb}
\usepackage{amsmath}
\usepackage{epstopdf}
\usepackage{hyperref}
\usepackage{xcolor}

\newcommand{\gradrad}{\ensuremath{\nabla_{\rm{rad}}}}
\newcommand{\gradad}{\ensuremath{\nabla_{\rm{ad}}}}
\newcommand{\justgrad}{\ensuremath{\nabla}}
\newcommand{\delp}{\ensuremath{\delta_{\rm{p}}}}
\newcommand{\Fbot}{\ensuremath{F_{\rm{bot}}}}
\newcommand{\Ftot}{\ensuremath{F_{\rm{tot}}}}
\newcommand{\Frad}{\ensuremath{F_{\rm{rad}}}}
\newcommand{\Fconv}{\ensuremath{F_{\rm{conv}}}}
\newcommand{\Fcz}{\ensuremath{F_{\rm{cz}}}}
\newcommand{\mP}{\ensuremath{\mathcal{P}}}

\newcommand{\Lcz}{\ensuremath{L_{\rm{CZ}}}}
\newcommand{\mR}{\ensuremath{\mathcal{R}}}
\newcommand{\mS}{\ensuremath{\mathcal{S}}}
\newcommand\Pran{\ensuremath{\mathrm{Pr}}}
\newcommand{\brunt}{Brunt-V\"{a}is\"{a}l\"{a}}

\newcommand{\angles}[1]{\langle #1 \rangle}

\renewcommand{\vec}[1]{\boldsymbol{#1}}
\newcommand{\M}[1]{\mathbf{#1}}
\renewcommand{\dot}{\vec{\cdot}}
\renewcommand{\bar}[1]{\overline{#1}}
\newcommand{\grad}{\vec{\nabla}}
\newcommand{\cross}{\vec{\times}}

\newcommand{\editone}[1]{#1}
\newcommand{\edittwo}[1]{#1}
\received{July 28, 2021}
\revised{October 19, 2021}
\accepted{}
\published{}
\submitjournal{ApJ}

\shorttitle{Convective Penetration}
\shortauthors{Anders et al}

\begin{document}

%%%% Title and Abstract
\title{\editone{Stellar convective penetration: parameterized theory and dynamical simulations}}
\author[0000-0002-3433-4733]{Evan H. Anders}
\affiliation{CIERA, Northwestern University, Evanston IL 60201, USA}
\author[0000-0001-5048-9973]{Adam S. Jermyn}
\affiliation{Center for Computational Astrophysics, Flatiron Institute, New York, NY 10010, USA}
\author[0000-0002-7635-9728]{Daniel Lecoanet}
\affiliation{CIERA, Northwestern University, Evanston IL 60201, USA}
\affiliation{Department of Engineering Sciences and Applied Mathematics, Northwestern University, Evanston IL 60208, USA}
\author[0000-0001-8935-219X]{Benjamin P. Brown}
\affiliation{Department Astrophysical and Planetary Sciences \& LASP, University of Colorado, Boulder, CO 80309, USA}

\correspondingauthor{Evan H. Anders}
\email{evan.anders@northwestern.edu}

\begin{abstract}
Most stars host convection zones in which heat is transported directly by fluid motion, but the behavior of convective boundaries is not well understood.
Here we present 3D numerical simulations which exhibit penetration zones: regions where the entire luminosity \emph{could} be carried by radiation, but where the temperature gradient is approximately adiabatic and convection is present.
To parameterize this effect, we define the ``penetration parameter'' $\mP$ which compares how far the radiative gradient deviates from the adiabatic gradient on either side of the Schwarzschild convective boundary.
Following \citet{roxburgh1989} and \citet{zahn1991}, we construct an energy-based theoretical model in which $\mP$ controls the extent of penetration.
We test this theory using 3D numerical simulations which employ a simplified Boussinesq model of stellar convection.
\edittwo{
    The convection is driven by internal heating and we use a height-dependent radiative conductivity; this allows us to separately specify $\mP$ and the stiffness $\mS$ of the radiative-convective boundary.
}We find significant convective penetration in all simulations.
Our simple theory describes the simulations well.
Penetration zones can take thousands of overturn times to develop, so long simulations or accelerated evolutionary techniques are required.
In stars, we expect $\mathcal{P} \approx 1$ and in this regime our results suggest that convection zones may extend beyond the Schwarzschild boundary by up to $\sim$20-30\% of a mixing length.
We present a MESA stellar model of the Sun which employs our parameterization of convective penetration as a proof of concept.
We discuss prospects for extending these results to more realistic stellar contexts.
\end{abstract}

%%%% Body of paper
\section{Introduction}
\label{sec:introduction}

\subsection{Context}
Convection is a crucial mechanism for transporting heat in stars~\citep{woosley_etal_2002, hansen_etal_2004, christensen-dalsgaard_2021}, and convective dynamics influence many poorly-understood stellar phenomena.
For example, convection drives the magnetic dynamo of the Sun, leading to a whole host of emergent phenomena collectively known as solar activity \citep{brun_browning_2017}.
Convection also mixes chemical elements in stars, which can modify observed surface abundances or inject additional fuel into their cores, thereby extending stellar lifetimes \citep{salaris_cassisi_2017}.
Furthermore, convective motions excite waves, which can be observed and used to constrain the thermodynamic structure of stars \citep{aerts2010, basu2016}.
A complete and nuanced understanding of convection is therefore crucial for understanding stellar structure and evolution, and for connecting this understand to observations.

Despite decades of study, robust parameterizations for the mechanisms broadly referred to as ``convective overshoot'' remain elusive, and improved parameterizations could resolve many discrepancies between observations and structure models.
In the stellar structure literature, ``convective overshoot'' refers to any convectively-driven mixing which occurs beyond the boundaries of the Ledoux-unstable zone.
This mixing can influence, for example, observed surface lithium abundances in the Sun and solar-type stars, which align poorly with theoretical predictions \citep{pinsonneault1997, carlos_etal_2019, dumont_etal_2021}.
Furthermore, modern spectroscopic observations suggest a lower solar metallicity than previously thought, and models computed with modern metallicity estimates and opacity tables have shallower convection zones than helioseismic observations suggest \citep{basu_antia_2004, bahcall_etal_2005, bergemann_serenelli_2014, vinyoles_etal_2017, asplund_etal_2021}; modeling and observational discrepancies can be reduced with additional mixing below the convective boundary \citep{christensen-dalsgaard_etal_2011}.

Beyond the Sun, overshooting in massive stars with convective cores must be finely tuned as a function of stellar mass, again pointing to missing physics in our current parameterizations \citep{claret_torres_2018, jermyn_etal_2018, viani_basu_2020, martinet_etal_2021, pedersen_etal_2021}.
Since core convective overshoot increases the reservoir of fuel available for nuclear fusion at each stage in stellar evolution, improved models of core convective boundary mixing could have profound impacts on the post-main sequence evolution and remnant formation of massive stars \citep{farmer_etal_2019, higgins_vink_2020}.

In order to ensure that models can be evolved on fast (human) timescales, 1D stellar evolution codes rely on simple parameterizations of convection \citep[e.g., mixing length theory,][]{bohm-vitense1958} and convective overshoot \citep{shaviv_salpeter_1973, maeder1975, herwig2000, paxton_etal_2011, paxton_etal_2013, paxton_etal_2018, paxton_etal_2019}.
While some preliminary work has been done to couple 3D dynamical convective simulations with 1D stellar evolution codes \citep{jorgensen_weiss_2019}, these calculations are prohibitively expensive to perform at every timestep in a stellar evolution simulation.
To resolve discrepancies between stellar evolution models and observations, a more complete and \emph{parameterizeable} understanding of convective overshoot is required.

The broad category of ``convective overshoot'' in the stellar literature is an umbrella term for a few hydrodynamical processes \citep{zahn1991, brummell_etal_2002, korre_etal_2019}.
Motions which extend beyond the convective boundary but do not adjust the thermodynamic profiles belong to a process called ``\emph{convective overshoot}'' in the fluid dynamics literature.
Convection zones can expand through a second process called ``\emph{entrainment},'' through which motions erode composition gradients or modify the radiative gradient \citep[][]{meakin_arnett_2007, viallet_etal_2013, cristini_etal_2017, jones_etal_2017, fuentes_cumming_2020, horst_etal_2021}.
The primary focus of this work is a third process called ``\emph{convective penetration}''.
Convective penetration occurs when motions mix the entropy gradient towards the adiabatic in a region that is stable by the Schwarzschild criterion.

\begin{figure*}[t]
\centering
\includegraphics[width=\textwidth]{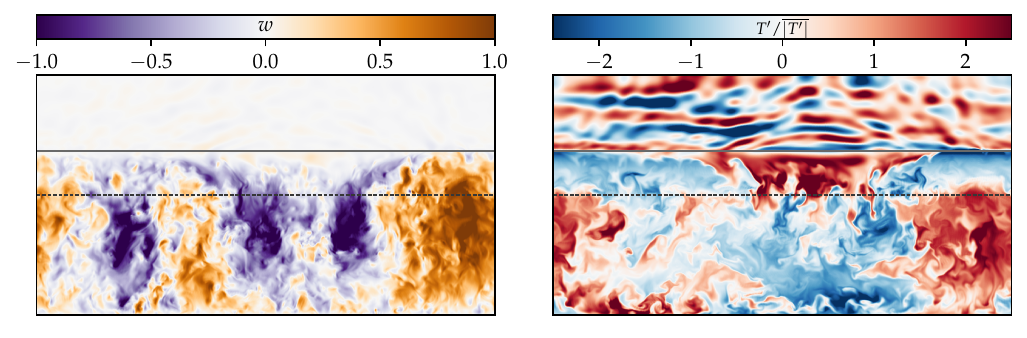}
\caption{
Vertical slice through a simulation with $\mR = 6.4 \times 10^3$, $\mP_D = 4$ and $\mS = 10^3$ (see Sec.~\ref{sec:simulation_details}).
The dashed horizontal line denotes the Schwarzschild convective boundary where $\gradad = \gradrad$.
The top of the penetrative zone ($\delta_{0.1}$, see Sec.~\ref{sec:simulation_details}) is shown by a solid horizontal line.
(Left) Vertical velocity is shown; orange convective upflows extend far past the Schwarzschild boundary of the convection zone but stop abruptly at the top of the penetration zone where $\justgrad$ departs from $\gradad$.
(Right) Temperature fluctuations, normalized by their average magnitude at each height to clearly display all dynamical features.
%Unlike the vertical velocity, $T'$ shows distinctly different behavior in the CZ and PZ, switching sign at the Schwarzschild boundary of the convection zone.
\label{fig:vertical_dynamics_panels}
}
\end{figure*}

Convective overshoot, entrainment, and penetration have been studied in the laboratory and through numerical simulations for decades, and the state of the field has been regularly reviewed \citep[e.g.,][]{marcus_etal_1983, zahn1991, browning_etal_2004, rogers_etal_2006, viallet_etal_2015, korre_etal_2019}.
Experiments exhibiting extensive expansion of convection zones via entrainment have a long history \citep[e.g.,][and this process is often confusingly called ``penetration'']{musman1968, deardorff_etal_1969, moore_weiss_1973}.
Modern numerical experiments often examine the importance of the ``stiffness'' $\mS$ of a radiative-convective interface.
$\mS$ compares the relative stability of a radiative zone and an adjacent convection zone according to some measure like a dynamical frequency or characteristic entropy gradient.
Some recent studies in simplified Boussinesq setups exhibit stiffness-dependent convection zone expansion via entrainment \citep{couston_etal_2017, toppaladoddi_wettlaufer_2018}; others find stiffness-dependent pure overshoot \citep{korre_etal_2019}.
A link between $\mS$ and the processes of entrainment and overshoot has seemingly emerged, but a mechanism for penetration remains elusive.

Many studies in both Cartesian and spherical geometries have exhibited hints of penetrative convection.
Some authors report clear mixing of the entropy gradient beyond the nominal convecting region \citep{hurlburt_etal_1994, saikia_etal_2000, brummell_etal_2002, rogers_glatzmaier_2005, rogers_etal_2006, kitiashvili_etal_2016}, but it is often unclear how much mixing is due to changes in the location of the Schwarzschild boundary (entrainment) and how much is pure penetration.
Other authors present simulations with dynamical or flux-based hints of penetration such as a negative convective flux or a radiative flux which exceeds the total system flux, but do not clearly report the value of the entropy gradient \citep{hurlburt_etal_1986, singh_etal_1995, browning_etal_2004, brun_etal_2017, pratt_etal_2017}.
Still other simulations show negligible penetration \citep[e.g.,][]{cai2020, higl_etal_2021}.
Even detailed studies which sought a relationship between penetration depth and stiffness $\mS$ have presented contradictory results.
Early work by e.g., \citet{hurlburt_etal_1994} and \citet{singh_etal_1995} hinted at a link between $\mS$ and penetration length, at least for low values of $\mS$.
Subsequent simulations by \citet{brummell_etal_2002} exhibit a weak scaling of penetration depth with $\mS$; the authors interpret this scaling as a sign of pure overshoot and claim their simulations do not achieve adiabatic convective penetration.
Still later simulations by \citet{rogers_glatzmaier_2005} demonstrate a negligible scaling of the penetration depth against $\mS$ at moderate values of $\mS$.
Prior simulations thus consistently show hints of penetration at low $\mS$ \citep[where results may not be relevant for stars,][]{couston_etal_2017}, but present confusing and contradictory results at moderate-to-high $\mS$.

There are hints in the literature that convective penetration may depend on energy fluxes.
\citet{roxburgh1978, roxburgh1989, roxburgh1992, roxburgh1998} derived an ``integral constraint'' from the energy equation and found that a spatial integral of the flux puts an upper limit on the size of a theoretical penetrative region.
\citet{zahn1991} theorized that convective penetration should depend only on how steeply the radiative temperature gradient varies at the convective boundary.
Following \citet{zahn1991}'s work, \citet{rempel2004} derived a semianalytic model and suggested that inconsistencies seen in simulations of penetrative dynamics can be explained by the magnitude of the fluxes or luminosities driving the simulations.
Indeed, some simulations have tested this idea, and found that penetration lengths depend strongly on the input flux \citep{singh_etal_1998, kapyla_etal_2007, tian_etal_2009, hotta2017, kapyla2019}.
Furthermore, in the limit of low stiffness, the simulations of \citet{hurlburt_etal_1994} and \citet{rogers_glatzmaier_2005} may agree with Zahn's theory (although at high stiffness they disagree).
In light of these results, and the possible importance of energy fluxes, Roxburgh's integral constraint and Zahn's theory deserve to be revisited.

$\,$
\subsection{Convective penetration \& this study's findings}

Convective penetration is the process by which convective motions extend beyond the Schwarzschild-stable boundary and mix the entropy gradient to be nearly adiabatic.
\begin{quote}
\emph{
    In this paper, we present simulations which exhibit convective penetration.
}
\end{quote}
This process is phenomenologically described in Sec.~\ref{sec:central_results}.
\editone{
    In this work, the convection zone lies beneath an adjacent stable layer and convection penetrates upwards; our results equally apply to the reversed problem.
}

In order to understand this phenomenon, we derive theoretical predictions for the size of the penetrative zone based on the ideas of \citet{roxburgh1989} and \citet{zahn1991}.
%We design and carry out numerical experiments to test the idea that the penetration depth depends on the shape of the radiative gradient at the convective boundary.
\begin{quote}
\emph{
We find that the extent of convective penetration depends strongly on the shape and magnitude of the radiative gradient near the convective boundary.
}
\end{quote}
Thus, the penetration length can be calculated using the radiative conductivity (or opacity) \emph{profile} near the convective boundary.
\edittwo{
    We present simulations of internally heated convection in which both the Schwarzschild boundary location and the extent of convective penetration depend primarily on the depth-dependent radiative conductivity.
}

We present these findings as follows.
In Sec.~\ref{sec:central_results}, we present the central finding of this work: penetration zones in nonlinear convective simulations.
In Sec.~\ref{sec:theory}, we describe the equations used and derive a parameterized theory of convective penetration.
In Sec.~\ref{sec:simulation_details}, we describe our simulation setup and parameters.
In Sec.~\ref{sec:results}, we present the results of these simulations, with a particular focus on the height of the penetrative regions.
In Sec.~\ref{sec:solar_model}, we create and discuss a stellar model in MESA which has convective penetration.
Finally, we discuss pathways for future work in Sec.~\ref{sec:discussion}.

\section{Central Result: Convective Penetration}
\label{sec:central_results}

In Fig.~\ref{fig:vertical_dynamics_panels}, we display a snapshot of dynamics in an evolved simulation which exhibits convective penetration.
The simulation domain is a 3D Cartesian box, and this figure shows a vertical slice through the center of the domain.
In the left panel, we display the vertical velocity.
We see that convective motions extend beyond the Schwarzschild boundary of the convection zone, which is denoted by a horizontal dashed grey line.
These motions stop at the top of a penetration zone, denoted by a solid horizontal line, where the temperature gradient departs from adiabatic towards the radiative gradient.
In the right panel, we display temperature perturbations away from the time-evolving mean temperature profile.
We see that warm upwellings in the Schwarzschild-unstable convection zone (below the dashed line) become cold upwellings in the penetration zone (above the dashed line), and these motions excite gravity waves in the stable radiative zone (above the solid line).

\begin{figure}[t]
\centering
\includegraphics[width=\columnwidth]{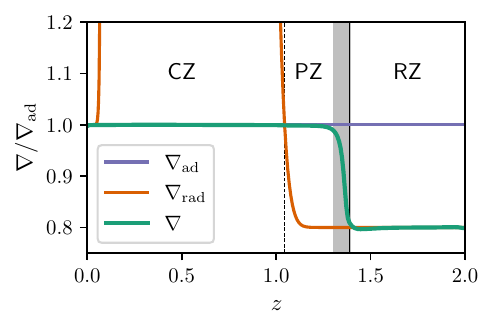}
\caption{
Horizontally- and temporally-averaged profiles of the thermodynamic gradients from the simulation in Fig.~\ref{fig:vertical_dynamics_panels}.
We plot $\justgrad$ (green) compared to $\gradad$ (purple, a constant) and $\gradrad$ (orange); note the extended penetration zone (PZ) where $\justgrad \approx \gradad > \gradrad$.
    \editone{
        The dashed vertical line denotes the Schwarzschild boundary of the convection zone (CZ), the solid vertical line denotes the bottom of the radiative zone (RZ), and the greyed region denotes the PZ-RZ boundary layer.
    }
\label{fig:grad_profiles}
}
\end{figure}

We further explore the simulation from Fig.~\ref{fig:vertical_dynamics_panels} in Fig.~\ref{fig:grad_profiles} by displaying time- and horizontally-averaged 1D profiles of the temperature gradient $\justgrad$ (defined in Sec.~\ref{sec:theory}).
The adiabatic gradient $\gradad$ (purple) has a constant value in the simulation.
Also shown is the radiative gradient $\gradrad$ (orange).
The domain exhibits a classical Schwarzshild-unstable convection zone (CZ) for $z \lesssim 1.04$ where $\gradrad > \gradad$; the upper boundary of this region is denoted by a dashed vertical line.
Above this point, $\gradrad < \gradad$ and the domain would be considered stable by the Schwarzschild criterion.
However, the evolved convective dynamics in Fig.~\ref{fig:vertical_dynamics_panels} have raised $\justgrad \rightarrow \gradad$ in an extended penetration zone (PZ) which extends from $1.04 \lesssim z \lesssim 1.3$.
\editone{
    Above $z \gtrsim 1.4$, $\justgrad \approx \gradrad$ in a classical stable radiative zone (RZ).
    Between $1.3 \lesssim z \lesssim 1.4$, there is a PZ-RZ boundary layer (referred to as the ``thermal adjustment layer'' in some prior studies) where convective motions give way to conductive transport and $\justgrad$ adjusts from $\gradad$ to $\gradrad$.
}

Our goals in this paper are to understand how these PZs form and to parameterize this effect so that it can be included in 1D stellar evolution calculations.

\section{Theory}
\label{sec:theory}
In this section we derive a theoretical model of convective penetration by examining the energetics and energy fluxes in the Schwarzschild-unstable convection zone (CZ) and penetration zone (PZ).
In Sec.~\ref{sec:theory_equations}, we describe our equations and problem setup and define the heat fluxes.
In Sec.~\ref{sec:theory_energy}, we build a parameterized theory based on the kinetic energy (KE) equation.
We find that imbalances in KE source terms within the CZ determine the extent of the PZ.
%We find that excess KE in the convection zone can cause a penetration zone to form.
By balancing the excess KE generation in the CZ with \editone{buoyant deceleration} and dissipation work terms in the PZ, we are able to derive the size of the PZ.
We find that a description of the size of a theoretical PZ does not depend on the often-considered stiffness, which measures the relative stability between the convection zone and an adjacent radiative zone.

\subsection{Equations \& flux definitions}
\label{sec:theory_equations}
Throughout this work, we will utilize a modified version of the incompressible Boussinesq equations,
\begin{align}
&\grad\dot\vec{u} = 0 
\label{eqn:incompressible} \\
&\partial_t \vec{u} + \vec{u}\dot\grad\vec{u} = -\frac{1}{\rho_0}\grad p + \frac{\rho_1}{\rho_0}\vec{g} + \nu\grad^2 \vec{u} 
\label{eqn:momentum} \\
&\partial_t T + \vec{u}\dot\grad T + w \gradad + \grad\dot[-k \grad \overline{T}] = \chi\grad^2 T' + Q
\label{eqn:temperature} \\
&\frac{\rho_1}{\rho_0} = -|\alpha| T.
\label{eqn:boussinesq}
\end{align}
Here, the density is decomposed into a uniform, constant background $\rho_0$ with fluctuations $\rho_1$ which appear only in the buoyancy force and depend on the temperature $T$ and the coefficient of thermal expansion $\alpha = \partial\ln\rho / \partial T$.
We define the velocity vector $\vec{u}$, \editone{the pressure $p$,} the viscous diffusivity $\nu$, the thermal diffusivity $\chi$, the bulk internal heating $Q$, the adiabatic gradient $\gradad$, and a height-dependent thermal conductivity\footnote{
    \edittwo{
        In a star, $\chi \equiv k$.
        We separate these values out of practicality, because simulations are well-resolved and numerically stable when $k \ll \chi$.
        The maximum vertical wavenumber of the $\bar{T}$ expansion is set by the stiffness (see Eqn.~\ref{eqn:stiffness}), not the radiative diffusivity $k$, so $k$ can be small.
        On the other hand, an expansion of the turbulent fluctuations $T'$ must include the cutoff wavenumber of the turbulent cascade, which is set by $\chi$.
        We separate $k$ and $\chi$ in order to explore simulations with a wider range of penetrative behaviors (per Eqn.~\ref{eqn:theory_P_defn}), as the theory presented here depends only on $k$.
        Note that as we increase the turbulence (the Reynolds number) in our simulations, we decrease $\chi$, and $\chi \rightarrow k$.
    }
} $k$.
We will consider Cartesian coordinates $(x, y, z)$ with a constant vertical gravity $\vec{g} = -g\hat{z}$.
Throughout this work, we will represent horizontal averages with bars ($\overline{\,\cdot\,}$) and fluctuations away from those averages with primes ($'$).
Thus, in Eqn.~\ref{eqn:temperature}, $\bar{T}$ is the horizontally averaged temperature and $T'$ are fluctuations away from that; both of these fields evolve in time according to Eqn.~\ref{eqn:temperature}.

Assuming convection reaches a time-stationary state, the heat fluxes are found by horizontally-averaging then vertically integrating Eqn.~\ref{eqn:temperature} to find
\begin{equation}
\overline{\Ftot} = \overline{\Frad} + \overline{\Fconv} = \int Q dz + \Fbot,
\label{eqn:flux_definition}
\end{equation}
where $\Fbot$ is the flux carried at the bottom of the domain, and $\overline{\Ftot}$ is the total flux, which can vary in height due to the heating $Q$.
The mean temperature profile $\overline{T}$ carries the radiative flux $\bar{\Frad} = -k \grad \overline{T}$.
We note that $k$ and $-\partial_z \bar{T}$ fully specify $\bar{\Frad}$ and in turn the convective flux, $\bar{\Fconv} = \bar{\Ftot} - \bar{\Frad}$.
We define the temperature gradient and radiative temperature gradient 
\begin{equation}
\justgrad \equiv -\partial_z \bar{T} \qquad
\gradrad \equiv \frac{\bar{\Ftot}}{k}.
\label{eqn:gradrad_definition}
\end{equation}
We have defined the $\justgrad$'s as positive quantities to align with stellar structure conventions and intuition.
Marginal stability is achieved when $\justgrad = \gradad$, which we take to be a constant.
We note that the classical Schwarzschild boundary of the convection zone is the height $z = L_s$ at which $\gradrad = \gradad$ and $\bar{\Fconv} = 0$.

The addition of a nonzero $\gradad$ to Eqn.~\ref{eqn:temperature} was derived by \citet{spiegel_veronis_1960} and utilized by e.g., \citet{korre_etal_2019}.
In this work, we have decomposed the radiative diffusivity into a background portion ($\grad\dot \bar{\Frad}$) and a fluctuating portion ($\chi \grad^2 T'$); by doing so, we have introduced a height-dependent $\gradrad$ to the equation set while preserving the diffusive behavior on fluctuations felt by classical Rayleigh-B\'{e}nard convection.
Here, we will assume a model in which an unstable convection zone ($\gradrad > \gradad$) sits below a stable radiative zone ($\gradrad < \gradad$), but in this incompressible model where there is no density stratification to break the symmetry of upflows and downflows, precisely the same arguments can be applied to the inverted problem.

\subsection{Kinetic energy \& the dissipation-flux link}
\label{sec:theory_energy}

\begin{figure}[t!]
\centering
\includegraphics[width=\columnwidth]{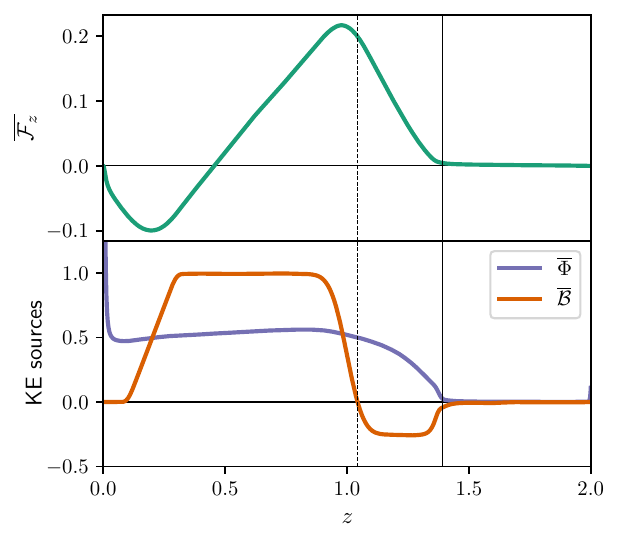}
\caption{
Temporally- and horizontally-averaged profiles from Eqn.~\ref{eqn:kinetic_energy_1D} in the simulation in Fig.~\ref{fig:vertical_dynamics_panels}.
The vertical dashed line denotes the Schwarzschild CZ boundary, and the vertical solid line corresponds to the top of the PZ.
(upper) Kinetic energy fluxes $\bar{\mathcal{F}_{\editone{z}}}$, which go to zero at the bottom boundary and the top of the PZ.
    (bottom) Source terms from Eqn.~\ref{eqn:kinetic_energy_1D} normalized by the maximum of $\overline{\mathcal{B}}$ ($\bar{\mathcal{F}_{\editone{z}}}$ in the upper panel is similarly normalized).
The buoyancy source $\bar{\mathcal{B}}$ changes sign at the Schwarzschild boundary, and $\bar{\Phi}$ is positive-definite.
\label{fig:theory_profiles}
}
\end{figure}

Taking a dot product of the velocity and Eqn.~\ref{eqn:momentum} reveals the kinetic energy equation,
\begin{equation}
\frac{\partial \mathcal{K}}{\partial t}
    + \grad\dot\editone{\vec{\mathcal{F}}}
= \mathcal{B} - \Phi,
%\frac{\partial}{\partial t}\left(\frac{|\vec{u}|^2}{2}\right) 
%+ \grad\dot\left[\vec{u}\left(\frac{|\vec{u}|^2}{2} + \frac{p}{\rho_0}\right) - \nu\vec{u}\cross\vec{\omega}\right]
%= |\alpha| g w T - \nu|\vec{\omega}|^2,
\label{eqn:kinetic_energy}
\end{equation}
where we define the kinetic energy $\mathcal{K} \equiv |\vec{u}|^2/2$, the fluxes of kinetic energy $\editone{\vec{\mathcal{F}}} \equiv \left[\vec{u}(\mathcal{K} + p/\rho_0) - \nu\vec{u}\cross\vec{\omega} \right]$, the buoyant energy generation rate $\mathcal{B} \equiv |\alpha| g w T'$, and the viscous dissipation rate $\Phi \equiv \nu |\vec{\omega}|^2$ where $\vec{\omega} = \grad\cross\vec{u}$ is the vorticity and $|\vec{u}|^2 = \vec{u}\dot\vec{u}$ \& $|\vec{\omega}|^2 = \vec{\omega}\dot\vec{\omega}$.
We next take a horizontal- and time-average of Eqn.~\ref{eqn:kinetic_energy} (we absorb the time-average into the horizontal-average $\bar{\,\cdot\,}$ notation for simplicity).
Assuming that $\bar{\mathcal{K}}$ reaches a statistically stationary state, convective motions satisfy
\begin{equation}
    \frac{d\bar{\mathcal{F}_{\editone{z}}}}{dz} = \bar{\mathcal{B}} - \bar{\Phi},
\label{eqn:kinetic_energy_1D}
\end{equation}
\editone{where $\mathcal{F}_z$ is the z-component of $\vec{\mathcal{F}}$.}
Each profile in Eqn.~\ref{eqn:kinetic_energy_1D} is shown in Fig.~\ref{fig:theory_profiles} for the simulation whose dynamics are displayed in Fig.~\ref{fig:vertical_dynamics_panels}.
As in Fig.~\ref{fig:grad_profiles}, the Schwarzschild CZ boundary is plotted as a dashed line, and the top of the PZ is plotted as a solid vertical line.
In the top panel, we display $\bar{\mathcal{F}_{\editone{z}}}$, neglecting the viscous flux term which is only nonzero in a small region above the bottom boundary.
We see that $\bar{\mathcal{F}_{\editone{z}}}$ is zero at the bottom boundary (left edge of plot) and at the top of the PZ.
In the bottom panel, we plot $\bar{\mathcal{B}}$ and $\bar{\Phi}$; we see that $\bar{\mathcal{B}}$ changes sign at the Schwarzschild CZ boundary, and that $\bar{\Phi}$ is positive-definite.

At the boundaries of the convecting region, $\bar{\mathcal{F}_{\editone{z}}}$ is zero (Fig.~\ref{fig:theory_profiles}, upper panel).
We integrate Eqn.~\ref{eqn:kinetic_energy_1D} vertically between these zeros to find
\begin{equation}
\int \bar{\mathcal{B}}\,dz = \int \bar{\Phi}\,dz.
\label{eqn:integral_constraint}
\end{equation}
Integral constraints of this form are the basis for a broad range of analyses in Boussinesq convection \citep[see e.g.,][]{ahlers_etal_2009, goluskin2016} and were considered in the context of penetrative stellar convection by \citet{roxburgh1989}.
Eqn.~\ref{eqn:integral_constraint} is the straightforward statement that work by buoyancy on large scales must be balanced by viscous dissipation on small scales.

We break up the convecting region into a Schwarzschild-unstable ``convection zone'' (CZ) and an extended ``penetration zone'' (PZ); we assume that convective motions efficiently mix $\justgrad \rightarrow \gradad$ in both the CZ and PZ.
The buoyant energy generation is proportional to the convective flux, $\bar{\mathcal{B}} = |\alpha|g\bar{w T'} = |\alpha| g \bar{\Fconv}$, and is positive in the CZ and negative in the PZ (see Fig.~\ref{fig:theory_profiles}, bottom panel).
Breaking up Eqn.~\ref{eqn:integral_constraint}, we see that
\begin{equation}
%\int_{\rm{CZ}} \bar{\mathcal{B}}\,dz \qquad=\qquad
\int_{\rm{CZ}} \bar{\mathcal{B}}\,dz \,\,=\,\,
\int_{\rm{CZ}} \bar{\Phi}\,dz + \int_{\rm{PZ}} \bar{\Phi}\,dz + \int_{\rm{PZ}}(-\bar{\mathcal{B}})\,dz.
\label{eqn:constraint_cz_pz_split}
\end{equation}
Eqn.~\ref{eqn:constraint_cz_pz_split} is arranged so that the (positive) buoyant engine of convection is on the left-hand side, and the (positive) sinks of work are on the RHS.
If viscous dissipation in the CZ does not balance the buoyant generation of energy in the CZ, the kinetic energy of the convective flows grows, resulting in a penetrative region.
This region grows with time until Eqn.~\ref{eqn:constraint_cz_pz_split} is satisfied.
We see that the viscous dissipation and \editone{buoyant deceleration} felt by flows in the PZ determine its size.
We now define
\begin{equation}
f \equiv \frac{\int_{\rm{CZ}} \bar{\Phi}\,dz}{\int_{\rm{CZ}}\bar{\mathcal{B}}\,dz},
\label{eqn:f_defn}
\end{equation}
the measurable fraction of the buoyant engine consumed by CZ dissipation.
Eqn.~\ref{eqn:constraint_cz_pz_split} can then be rewritten as
\begin{equation}
\frac{\int_{\rm{PZ}}(-\bar{\mathcal{B}})\,dz}{\int_{\rm{CZ}} \bar{\mathcal{B}}\,dz}
+ \frac{\int_{\rm{PZ}} \bar{\Phi}\,dz }{\int_{\rm{CZ}} \bar{\mathcal{B}}\,dz}
= (1 - f).
\label{eqn:first_pz_parameterization}
\end{equation}
We will measure and report the values of $f$ achieved in our simulations in this work.
Eqn.~\ref{eqn:first_pz_parameterization} provides two limits on a hypothetical PZ:
\begin{enumerate}
\item In the limit that $f \rightarrow 0$, viscous dissipation is inefficient.
Reasonably if we also assume that $\int_{\rm{PZ}}\bar{\Phi}\,dz \rightarrow 0$, Eqn.~\ref{eqn:first_pz_parameterization} states that the PZ must be so large that its negative buoyant work is equal in magnitude to the positive buoyant work of the CZ.
This is the integral constraint on the maximum size of the PZ that \citet{roxburgh1989} derived.
\item In the limit that $f \rightarrow 1$, viscous dissipation efficiently counteracts the buoyancy work in the CZ.
Per Eqn.~\ref{eqn:first_pz_parameterization}, the positive-definite PZ terms must approach zero and no PZ develops in this limit.
This is mathematically equivalent to standard boundary-driven convection experiments.
\end{enumerate}
In general, we anticipate from the results of e.g., \citet{currie_browning_2017} that $f$ is closer to 1 than 0, but its precise value must be measured from simulations.
Indeed, we find that $f \gg 0$ but $f < 1$ in our simulations (see e.g., Fig.~\ref{fig:theory_profiles}, bottom panel\footnote{the bulk dynamics suggest by eye $f \sim 0.5$, but due to e.g., the height dependence of $\bar{\mathcal{B}}$ in our simulations we measure $f \approx 0.74$.}).
Our simulations produce typical values of $f \sim 0.7$.

Assuming that a PZ of height $\delp$ develops above a CZ of depth $L_{\rm{CZ}}$, we model the PZ dissipation as
\begin{equation}
\int_{\rm{PZ}} \bar{\Phi}\,dz = \xi\frac{\delp}{L_{\rm{CZ}}}\int_{\rm{CZ}}\bar{\Phi}\,dz = \xi \delp \Phi_{\rm{CZ}}.
\label{eqn:xi_defn}
\end{equation}
Here $\Phi_{\rm{CZ}}$ is the volume-averaged dissipation rate in the CZ and $\xi$ is a measurable parameter in ${[0, 1]}$ that describes the shape of the dissipation profile as a function of height in the PZ.
In words, we assume that $\bar{\Phi}(z = L_s) \approx \Phi_{\rm{CZ}}$ at the CZ-PZ boundary and that $\bar{\Phi}$ decreases with height in the PZ.
The shape of $\bar{\Phi}$ determines $\xi$; a linear falloff gives $\xi = 1/2$, a quadratic falloff gives $\xi = 2/3$, and $\xi = 1$ assumes no falloff.
With this parameterization, and $\bar{\mathcal{B}} \propto \bar{\Fconv}$, we rewrite Eqn.~\ref{eqn:first_pz_parameterization},
\begin{equation}
-\frac{\int_{\rm{PZ}}\bar{\Fconv}\,dz}{\int_{\rm{CZ}}\bar{\Fconv}\,dz} + f\xi\frac{\delp}{L_{\rm{CZ}}}
= (1 - f).
\label{eqn:theory_fraction}
\end{equation}
The fundamental result of this theory is Eqn.~\ref{eqn:theory_fraction}, which is a parameterized and generalized form of \citet{roxburgh1989}'s integral constraint.
This equation is also reminiscent of \citet{zahn1991}'s theory, and says that the size of a PZ is set by the profile of $\gradrad$ near the convective boundary.
A parameterization like Eqn.~\ref{eqn:theory_fraction} can be implemented in stellar structure codes and used to find the extent of penetration zones under the specification of $f$ and $\xi$.
\editone{
    We note that an implementation of Eqn.~\ref{eqn:theory_fraction} likely requires an \emph{iterative} solve, as the penetration zone depth ($\delp$) and thus the PZ integral of the flux, are not known \emph{a-priori}.
}
The parameters $f$ and $\xi$ are measurables which can be constrained by direct numerical simulations, and we will measure their values in this work.
In general, we expect that $f$ and $\xi$ should not change too drastically with other simulation parameters.

In order to derive a specific prediction for the PZ height, one must specify the vertical shape of $\overline{\Fconv}$.
We will study two cases in this work, laid out below.
In both of these cases, we define a nondimensional ``Penetration Parameter'' whose magnitude is set by the ratio of the convective flux slightly above and below the Schwarzschild convective boundary $L_s$ (assuming $\justgrad = \gradad$ in the CZ and PZ),
\begin{equation}
%\mP \equiv -\frac{\overline{\Fconv}(z = L_s - \epsilon)}{\overline{\Fconv}(z = L_s + \epsilon)}
\mP \equiv -\frac{\bar{\Fconv}_{\rm{CZ}}}{\bar{\Fconv}_{\rm{PZ}}}.
\label{eqn:theory_P_defn}
\end{equation}
Since $\Fconv < 0$ in the PZ, the sign of $\mP$ is positive.
Intuitively, $\mP$ describes which terms are important in Eqn.~\ref{eqn:first_pz_parameterization}.
When $\mP \ll 1$, the buoyancy term dominates in the PZ and dissipation can be neglected there.
When $\mP \gg 1$, buoyancy is negligible and dissipation constrains the size of the PZ.
When $\mP \sim 1$, both terms matter.
\editone{
    In this work, we have assumed that $\mP$ and $\xi$ are fully independent parameters.
    We make this choice because $\mP$ can be determined directly from a known conductivity profile or stratification, whereas $\xi$ is a measurable of evolved nonlinear convective dynamics.
    However, it is possible that there is an implicit relationship between these parameters (as $\mP$ increases, so too does the extent of the PZ, which likely in turn modifies the value of $\xi$).
}

\subsubsection{Case I: Discontinuous flux}
\label{sec:discontinuous_theory}
We first consider a model which satisfies
\begin{equation}
\overline{\Fconv}(z) = \Fcz \begin{cases}
1			&	z \leq L_s,\\
-\mP_D^{-1}  & 	z > L_s 
\end{cases}.
\end{equation}
Here, $\Fcz$ is a constant value of flux carried in the convection zone and $\mP_D$ is the penetration parameter (subscript D for discontinuous case).
Plugging this functional form of the flux into Eqn.~\ref{eqn:theory_fraction}, and integrating the CZ over a depth $L_{\rm{CZ}}$ below $L_s$ and the PZ over a height $\delp$ above $L_s$, we predict
\begin{equation}
\frac{\delp}{\Lcz} = \mP_D \frac{1 - f}{1 + \xi f \mP_D}.
\label{eqn:discontinuous_prediction}
\end{equation}
Assuming that $f$ and $\xi$ are weak functions of $\mP_D$, we see that, for small $\mP_D$, the size of the penetration region is linearly proportional to $\mP_D$, but saturates as $\mP_D \rightarrow \infty$ due to dissipation.
Intuitively, this result makes sense: as $\mP_D$ grows, the magnitude of $\overline{\Fconv}$ and the \editone{deceleration caused by} buoyancy in the PZ shrink, resulting in larger penetrative regions (but this growth cannot extend indefinitely).

\subsubsection{Case II: Piecewise linear flux}
\label{sec:linear_theory}
We next examine a model where the derivative of $\overline{\Fconv}(z)$ may be discontinuous at the CZ-PZ boundary,
\begin{equation}
\overline{\Fconv}(z) = 
\frac{\partial \Frad}{\partial z}\bigg|_{\rm{CZ}}
\begin{cases}
(L_s - z) & z \leq L_s \\
-\mP_L^{-1} (z - L_s) & z > L_s
\end{cases},
\label{eqn:linear_flux_theory}
\end{equation}
where $(\partial \Frad / \partial z)|_{\rm{CZ}}$ is a constant and $\mP_L$ is the penetration parameter (subscript L for linear case).
When $\mP_L = 1$, $\bar{\Fconv}$ is a linear profile that crosses through zero at $z = L_s$.
Solving Eqn.~\ref{eqn:theory_fraction} with Eqn.~\ref{eqn:linear_flux_theory} and integrating over $L_{\rm{CZ}}$ in the CZ and $\delp$ in the PZ, we retrieve a quadratic equation.
This equation has two solution branches, only one of which corresponds to a positive value of $\delp$.
On that branch, we find
\begin{equation}
\frac{\delp}{\Lcz} = \sqrt{\mP_L (1 - f)} \,\,(\sqrt{\zeta^2 + 1} - \zeta),
\label{eqn:linear_prediction}
\end{equation}
where $\zeta \equiv (\xi f/2)\sqrt{\mP_L/(1-f)}$.
We expect the penetration height to be proportional to $\sqrt{\mP_L}$ for small values of $\mP_L$, and to again saturate at large values of $\mP_L$ \editone{(as $\mP_L \rightarrow \infty$, so too $\zeta \rightarrow \infty$, and $(\sqrt{\zeta^2 + 1} - \zeta) \rightarrow 0$).}

In this work, we will test Eqn.~\ref{eqn:theory_fraction} through the predictions of Eqns.~\ref{eqn:discontinuous_prediction} and \ref{eqn:linear_prediction}.
Our goals are to see if the predicted scalings with the penetration parameter $\mP$ are realized in simulations, and to measure the values of $f$ and $\xi$.

\section{Simulation Details}
We will now describe a set of simulations that test the predictions in Sec.~\ref{sec:theory}.
While many simulations of convection interacting with radiative zones have been performed by previous authors, ours differ in two crucial ways.
First, we construct our experiments so that $\mP$ and $\mS$ can be varied separately \editone{by driving convection with internal heating, thus avoiding strongly superadiabatic boundary layers where $\justgrad \rightarrow \gradrad$}.
$\mP$ is the ``Penetration Parameter,'' defined in Eqn.~\ref{eqn:theory_P_defn}, which compares the magnitude of the convective flux in the CZ and PZ; $\mS$ is the ``stiffness,'' defined in Eqn.~\ref{eqn:stiffness}, and compares the buoyancy frequency in the stable radiative zone to the convective frequency. 
We suspect that some past experiments have implicitly set $\mP \approx \mS^{-1}$, which would result in negligible penetration for high stiffness \editone{(see discussion following Eqn.~\ref{eqn:stiffness})}.
Second, as we will show in Sec.~\ref{sec:results}, the development of penetrative zones is a slow process and many prior studies did not evolve simulations for long enough to see these regions grow and saturate.

\editone{
    Appealing to the Buckingham $\pi$ theorem \citep{buckingham_1914}, we count nine fundamental input parameters in Eqns.~\ref{eqn:incompressible}-\ref{eqn:boussinesq}: $\rho_0$, $\alpha g$, $L_s$, $\nu$, $\chi$, $Q$, $\gradad$, $k_{\rm{CZ}}$, and $k_{\rm{RZ}}$.
    There are four fundamental dimensions (mass, length, time, and temperature), and so we are left with five independent prognostic parameters in setting up our system.
    For two of these parameters, we will choose the freefall Reynolds number and the Prandtl number, which are analagous to the Rayleigh and Prandtl numbers in Rayleigh-B\'{e}nard convection.
    The remaining three parameters are $\mS$, $\mP$, and an additional parameter $\mu$, which we will hold constant and which sets the ratio between $\gradrad$ and $\gradad$ in the convection zone.
}

We nondimensionalize Eqns.~\ref{eqn:incompressible}-\ref{eqn:boussinesq} on the length scale of the Schwarzschild-unstable convection zone $L_s$, the timescale of freefall across that convection zone 

\begin{equation}
    \editone{
    \tau_{\rm{ff}} = \left(\frac{L_s}{|\alpha| g Q_0}\right)^{1/3},
}
\end{equation}
and the temperature scale of the internal heating over that freefall time $\Delta T$; \editone{mass is nondimensionalized so that the freefall ram pressure $\rho_0(L_s/\tau_{\rm{ff}})^2 = 1$,}
\begin{equation}
\begin{split}
&T^* = (\Delta T)T = Q_0 \tau_{\rm{ff}} T,\qquad
Q^* = Q_0 Q,
\\
&\partial_{t^*} = \tau_{\rm{ff}}^{-1}\partial_t,\qquad\qquad\qquad\,\,\,
\grad^* = L_s^{-1} \grad,\,\,\,\,\,
\\
    &\vec{u}^* = u_{\rm{ff}}\vec{u} = \frac{L_s}{\tau_{\rm{ff}}} \vec{u}, \qquad\qquad
p^* = \rho_0 u_{\rm{ff}}^2\varpi,
\\
&k^* = (L_s^2 \tau_{\rm{ff}}^{-1})k,\qquad
\mR = \frac{u_{\rm{ff}} L_s}{\nu},\qquad
\Pran = \frac{\nu}{\chi}.
\end{split}
\end{equation}
For convenience, here we define quantities with $*$ (e.g., $T^*$) as being the ``dimensionful'' quantities of Eqns.~\ref{eqn:incompressible}-\ref{eqn:boussinesq}.
Henceforth, quantities without $*$ (e.g., $T$) are dimensionless.
The dimensionless equations of motion are
\label{sec:simulation_details}
\begin{align}
&\grad\dot\vec{u} = 0 
\label{eqn:nondim_incompressible} \\
&\partial_t \vec{u} + \vec{u}\dot\grad\vec{u} = -\grad \varpi + T \hat{z} + \mR^{-1}\grad^2 \vec{u}
\label{eqn:nondim_momentum} \\
\begin{split}
\partial_t T + \vec{u}\dot\grad T + w \grad_{\rm{ad}}  + \grad\dot[-k \grad \overline{T}]\qquad\qquad 
\\
\qquad\qquad\qquad\qquad\qquad= (\Pran\mR)^{-1}\grad^2 T' + Q.
\label{eqn:nondim_temperature}
\end{split}
\end{align}
We construct a domain in the range $z \in [0, L_z]$ and choose $L_z \geq 2$ so that the domain is at least twice as deep as the Schwarzschild-unstable convection zone.
We decompose the temperature field into a time-stationary initial background profile and fluctuations, $T(x, y, z, t) = T_0(z) + T_1(x, y, z, t)$.
$T_0$ is constructed with $\justgrad = \gradad$ for $z \leq L_s$, and $\justgrad = \gradrad$ above $z > L_s$.
We impose a fixed-flux boundary at the bottom of the box ($\partial_z T_1 = 0$ at $z = 0$) and a fixed temperature boundary at the top of the domain ($T_1 = 0$ at $z = L_z$).
We generally impose impenetrable, no-slip boundary conditions at the top and bottom of the box so that $\vec{u} = 0$ at $z = [0, L_z]$.
For a select few simulations, we impose stress-free instead of no-slip boundary conditions ($w = 0$ and $\partial_z u = \partial_z v = 0$ at $z = [0, L_z]$).

We impose a constant internal heating which spans only part of the convection zone,
\begin{equation}
Q = \begin{cases}
0		& z < 0.1\,\,\rm{or}\,\,z\geq 0.1 + \Delta_H,\\
Q_{\rm{mag}}		& 0.1 \leq z \leq 0.1 + \Delta_H
\end{cases}.
\label{eqn:sim_Q}
\end{equation}
The integrated flux through the system from heating is $F_H = \int_0^{L_z} Q_{\rm{mag}} dz = Q_{\rm{mag}}\Delta_H$.
Throughout this work we choose $Q_{\rm{mag}} = 1$ and $\Delta_H = 0.2$ so $F_H = 0.2$.
We offset this heating from the bottom boundary to $z = 0.1$ to avoid heating within the bottom impenetrable boundary layer where velocities go to zero and $k$ is small; this prevents strong temperature gradients from establishing there.
Furthermore, since the conductivity is not zero at the bottom boundary, the adiabatic temperature gradient there carries some flux.
\editone{
We specify the flux using
\begin{equation}
    \mu \equiv \frac{\Fbot}{F_H}
    \label{eqn:mu_defn}
\end{equation}
}
and we choose $\mu = 10^{-3}$ so that most of the flux in the convection zone is carried by the convection.

\editone{
    Throughout this paper, we assume that the convection zone is roughly adiabatically stratified.
    We therefore define a dynamical measure of the stiffness, rather than one based on e.g., the superadiabaticity of $\gradrad$ in the convection zone.
}
The average convective velocity depends on the magnitude of the convective flux, $\angles{\vec{|u|}} \approx F_H^{1/3} = (Q_{\rm{mag}}\Delta_H)^{1/3}$.
The characteristic convective frequency is ${f_{\rm{conv}} = \angles{\vec{|u|}} / L_s}$.
Empirically we find that for our choice of parameters, $\angles{\vec{|u|}} \approx 1$, so going forward we define $f_{\rm{conv}} = 1$.
The stiffness is defined,
\begin{equation}
\mS \equiv \frac{N^2}{f_{\rm{conv}}^2} = N^2,
\label{eqn:stiffness}
\end{equation}
where $N^2$ is the \brunt$\,$frequency in the radiative zone.
In our nondimensionalization, $N^2 = \gradad - \gradrad$ in the radiative zone.
We use $\mS$ as a control parameter.

\editone{
    In many prior studies, the stiffness has been set by the ratio of the subadiabaticity of $\gradrad$ in the RZ to the superadiabaticity of $\gradrad$ in the CZ,
    \begin{equation}
        \tilde{\mS} = \frac{|\gradrad - \gradad|_{\rm{RZ}}}{|\gradrad - \gradad|_{\rm{CZ}}} = \frac{N^2}{|\gradrad - \gradad|_{\rm{CZ}}}.
    \end{equation}
    In those studies, $\tilde{\mS}$ primarily describes the stratification of the initial state, but it also describes the stratification in superadiabatic boundary layers which drive convection.
    In this work, we maintain a nearly adiabatic convection zone without strongly superadiabatic regions by driving convection with an internal heating function which is offset from the lower boundary.
}

\editone{
    Previous work has not defined $\mP$, but its definition in our current study should apply to previous studies,
    \begin{equation}
        \mP = -\frac{k_{\rm{CZ}}(\gradrad - \gradad)_{\rm{CZ}}}{k_{\rm{RZ}}(\gradrad - \gradad)_{\rm{RZ}}}.
    \end{equation}
    We note that $\mP$ can be related to $\mS$ and $\tilde{\mS}$, $\mP = (k_{\rm{CZ}}/k_{\rm{RZ}}) \tilde{\mS}^{-1} = (k_{\rm{CZ}}/k_{\rm{RZ}}) (\gradrad - \gradad)_{\rm{CZ}} \mS^{-1}$.
    Our use of internal heating to decouple convective perturbations from $\gradrad$ in the CZ allows us to separately specify these nondimensional parameters.
    The distinction between $\mS$ and $\mP$ is perhaps clearer in the language of stellar evolution, where $\mS$ is roughly the inverse square Mach number of the convection while $\mP$ is set by the ratio of $\gradrad$ and $\gradad$.
}

Aside from $\mS$, $\mP$, and $\mu$, the two remaining control parameters $\mR$ and $\Pran$ determine the properties of the turbulence.
The value of $\mR$ corresponds to the value of the Reynolds number $\rm{Re} = \mR |\vec{u}|$, and we will vary $\mR$.
Astrophysical convection exists in the limit of $\Pran \ll 1$ \citep{garaud2021}; in this work we choose a modest value of $\Pran = 0.5$ which slightly separates the thermal and viscous scales while still allowing us to achieve convection with large Reynolds and P\'{e}clet numbers.

We now describe the two types of simulations conducted in this work (Case I and Case II).
We provide Fig.~\ref{fig:parameter_space} to visualize the portion of the parameter space that we have studied.
We denote two ``landmark cases'' using a purple box (Case I landmark) and an orange box (Case II landmark).
These landmark cases will be mentioned throughout this work.

\subsection{Case I: Discontinuous flux}
\label{sec:numerics_case1}
Most of the simulations in this paper have a discontinuous convective flux at the Schwarzschild convective boundary.
We achieve this by constructing a discontinuous radiative conductivity,
\begin{equation}
k = \begin{cases}
k_{\rm{CZ}}	&	z < 1 \\
k_{\rm{RZ}} &	z \geq 1
\end{cases},
\label{eqn:sim_discontinuous_k}
\end{equation}
where CZ refers to the convection zone and RZ refers to the radiative zone (some of which will be occupied by the penetrative zone PZ).
Using $\mS$ and $\mP_D$ as inputs and specifying the radiative flux at the bottom boundary and in the RZ defines this system,
\begin{equation}
\begin{split}
&k_{\rm{RZ}} = \frac{F_H}{f_{\rm{conv}}^2\mS\mP_D},\\
&k_{\rm{CZ}} = k_{\rm{RZ}}\frac{\mu}{1 + \mu + \mP_D^{-1}},\\
&\gradad = f_{\rm{conv}}^{2}\mS\mP_D(1 + \mu + \mP_D^{-1}),\\
&\gradrad = \gradad - f_{\rm{conv}}^{2}\mS.
\label{eqn:discontinuous_constants}
\end{split}
\end{equation}
\edittwo{
     Eqns.~\ref{eqn:discontinuous_constants} are found by solving the system of equations $\mS = (\gradad - \gradrad)/f_{\rm{conv}}^2$, $\mP_D = F_H/(k_{\rm{RZ}}[\gradad - \gradrad])$, $F_{\rm{bot}} = k_{\rm{CZ}}\gradad$, and ${F_{\rm{bot}} + F_H = k_{\rm{RZ}}\gradrad}$.
}

\begin{figure}[t]
\centering
\includegraphics[width=\columnwidth]{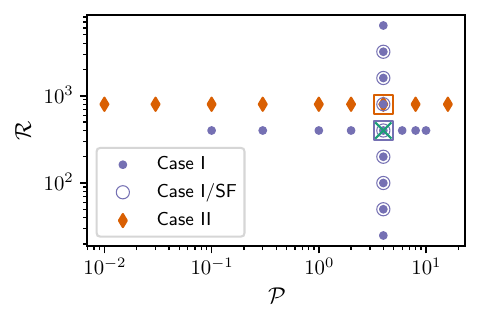}
\caption{
Each marker denotes a simulation conducted in this work in the $\mR-\mP$ parameter space at $\mS = 10^3$. 
Purple circles represent Case I (Sec.~\ref{sec:numerics_case1}) simulations and orange diamonds represent Case II (Sec.~\ref{sec:numerics_case2}) simulations; empty circular markers have stress-free (SF) boundary conditions and all other simulations have no-slip boundaries.
The green ``x'' at $\mP = 4$ and $\mR = 400$ denotes the location in $\mR-\mP$ parameter space where we vary $\mS$ in select Case I simulations.
Boxes denote the two ``landmark'' simulations.
The landmark Case I simulation has $\mR = 400$ and $\mP_D = 4$.
The landmark Case II simulation has $\mR = 800$ and $\mP_L = 4$.
Both landmark simulations have $\mS = 10^3$ and no-slip boundary conditions.
\label{fig:parameter_space}
}
\end{figure}

We study a sweep through each of the ($\mP_D$, $\mS$, $\mR$) parameter spaces while holding all other parameters constant (see Fig.~\ref{fig:parameter_space}).
We study an additional sweep through $\mR$ parameter space using stress-free boundaries to compare to our no-slip cases.
According to Eqn.~\ref{eqn:discontinuous_prediction}, we expect $\delp \propto \mP_D$.

\subsection{Case II: Piecewise linear flux}
\label{sec:numerics_case2}

We also study simulations where the flux's gradient may be discontinuous at the Schwarzschild convective boundary.
We achieve this by constructing a radiative conductivity with a piecewise discontinuous gradient,
\begin{equation}
\partial_z k = \partial_z k_0
\begin{cases}
1	&	z < 1 \\
\mP_L^{-1} &	z \geq 1
\end{cases}
\label{eqn:sim_linear_k}
\end{equation}
Since $k$ varies with height, formally the values of $\mS$ and $\mP$ also vary with height; we specify their values at $z = 2$.
By this choice, we require
\begin{equation}
\partial_z k_0 = \frac{F_{H}}{f_{\rm{conv}}^2L_s \mS \psi},\,\,
k_b = \frac{F_H \mu}{f_{\rm{conv}}^2\mS\psi},\,\,
\gradad = f_{\rm{conv}}^2 \mS \psi,
    \label{eqns:linear_constants}
\end{equation}
where $\psi \equiv 1 + \mP_L(1 + \mu)$.
We will study one sweep through $\mP_L$ space at fixed $\mR$ and $\mS$ (see Fig.~\ref{fig:parameter_space}).
According to Eqn.~\ref{eqn:linear_prediction}, we expect $\delta_p \propto \mP_L^{1/2}$.
\edittwo{
    We arrive at Eqns.~\ref{eqns:linear_constants} by solving the system of equations where ${F_{\rm{bot}} = k_{\rm{bot}}\gradad}$, ${F_{\rm{bot}} + F_H = k_{\rm{ad}}\gradad}$, ${k_{\rm{ad}} = k_{\rm{bot}} + \partial_z k_0 L_s}$, ${\mS = (\gradad - \justgrad_{\mathrm{rad},z=2L_s})/f_{\rm{conv}}^2}$, and ${\gradrad = F_{\rm{tot}}/k(z)}$.
}

\subsection{Numerics}
\label{sct:numerics}
We time-evolve equations \ref{eqn:nondim_incompressible}-\ref{eqn:nondim_temperature} using the Dedalus pseudospectral solver \citep{burns_etal_2020}\footnote{we use commit \href{https://github.com/DedalusProject/dedalus/commit/efb13bdaa09816dde3eee897bc2a15fc284ea2f1}{efb13bd}; the closest stable release to this commit is \href{https://github.com/DedalusProject/dedalus/releases/tag/v2.2006}{v2.2006}.} using timestepper SBDF2 \citep{wang&ruuth2008} and safety factor 0.35.
All fields are represented as spectral expansions of $n_z$ Chebyshev coefficients in the vertical ($z$) direction and as ($n_x$,$n_y$) Fourier coefficients in the horizontal ($x$,$y$) directions; our domains are therefore horizontally periodic.
We use a domain aspect ratio of two so that $x \in [0, L_x]$ and $y \in [0, L_y]$ with $L_x = L_y = 2 L_z$.
To avoid aliasing errors, we use the 3/2-dealiasing rule in all directions.
To start our simulations, we add random noise temperature perturbations with a magnitude of $10^{-3}$ to a background temperature profile $\overline{T}$; we discuss the choice of $\overline{T}$ in appendix \ref{app:accelerated_evolution}.
In some simulations we start with $\bar{T} = T_0$, described above, and in others we impose an established penetrative zone in the initial state $\bar{T}$ according to Eqn.~\ref{eqn:initial_grad}.

Spectral methods with finite coefficient expansions cannot capture true discontinuities.
In order to approximate discontinuous functions such as Eqns.~\ref{eqn:sim_Q}, \ref{eqn:sim_discontinuous_k}, and \ref{eqn:sim_linear_k}, we must use smooth transitions.
We therefore define a smooth Heaviside step function,
\begin{equation}
H(z; z_0, d_w) = \frac{1}{2}\left(1 + \mathrm{erf}\left[\frac{z - z_0}{d_w}\right]\right).
\label{eqn:heaviside}
\end{equation}
where erf is the error function.
In the limit that $d_w \rightarrow 0$, this function behaves identically to the classical Heaviside function centered at $z_0$.
For Eqn.~\ref{eqn:sim_Q} and Eqn.~\ref{eqn:sim_linear_k}, we use $d_w = 0.02$; while for Eqn.~\ref{eqn:sim_discontinuous_k} we use $d_w = 0.075$.
In all other cases, we use $d_w = 0.05$.

A table describing all of the simulations presented in this work can be found in Appendix~\ref{app:simulation_table}.
We produce the figures in this paper using matplotlib \citep{hunter2007, mpl3.3.4}.
All of the Python scripts used to run the simulations in this paper and to create the figures in this paper are publicly available in a git repository\footnote{\url{https://github.com/evanhanders/convective_penetration_paper}}, and in a Zenodo repository \citep{supp}.

\subsection{Penetration height measurements}
In our evolved simulations, the penetrative region has a nearly adiabatic stratification $\justgrad \approx \gradad$.
To characterize the height of the penetrative region, we measure how drastically $\justgrad$ has departed from $\gradad$.
We define the difference between the adiabatic and radiative gradient,
\begin{equation}
\Delta \equiv \gradad - \gradrad(z).
\end{equation}
We measure penetration heights in terms of ``departure points,'' or heights at which the realized temperature gradient $\justgrad$ has evolved away from the adiabatic $\gradad$ by some fraction $h < 1$ of $\Delta$.
Specifically,
\begin{equation}
L_s + \delta_{h} = \mathrm{max}(z) \,\,\mid\,\, \justgrad > (\gradad - h\,\Delta).
\label{eqn:delta_p_measures}
\end{equation}
In this work, we measure the 10\% ($\delta_{0.1}$, $h=0.1$), 50\% ($\delta_{0.5}$, $h=0.5$), and 90\% ($\delta_{0.9}$, $h=0.9$) departure points.
Using \citet{zahn1991}'s terminology, $\delta_{0.5}$ is the mean value of the top of the PZ while $\delta_{0.9} - \delta_{0.1}$ represents the width of the \editone{PZ-RZ boundary layer}. 
We find that these measurements based on the (slowly-evolving) thermodynamic profile provide a robust and straightforward measurement of penetration height \citep[for a discussion of alternate measurement choices, see][]{pratt_etal_2017}.

\begin{figure}[t]
\centering
\includegraphics[width=\columnwidth]{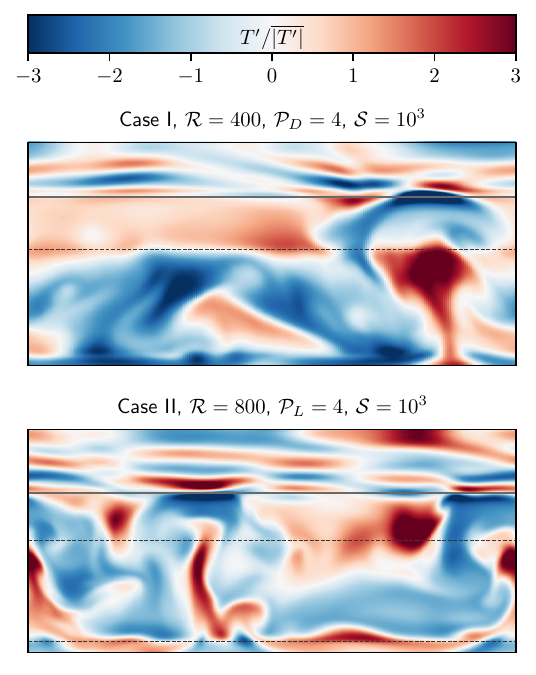}
\caption{
Temperature anomalies in vertical slices through the landmark simulations.
(top) Case I landmark ($\mR = 400$, $\mP_D = 4$, $\mS = 10^3$) and (bottom) Case II landmark ($\mR = 800$, $\mP_L = 4$, $\mS = 10^3$).
The temporally- and volume- averaged Reynolds number in the CZ is $\rm{Re} \sim 250$ in the top panel and $\rm{Re} \sim 350$ in the bottom panel.
A dashed horizontal line denotes the Schwarzschild convective boundary.
A solid line denotes the boundary between the penetrative and radiative zones.
The Case II simulation has an additional Schwarzschild boundary near the bottom of the domain due to the conductivity linearly increasing below the internal heating layer.
As in Fig.~\ref{fig:vertical_dynamics_panels}, temperature anomalies have different signs in the bulk CZ and PZ.
\label{fig:lessturb_dynamics_panels}
}
\end{figure}

\begin{figure*}[t]
\centering
\includegraphics[width=\textwidth]{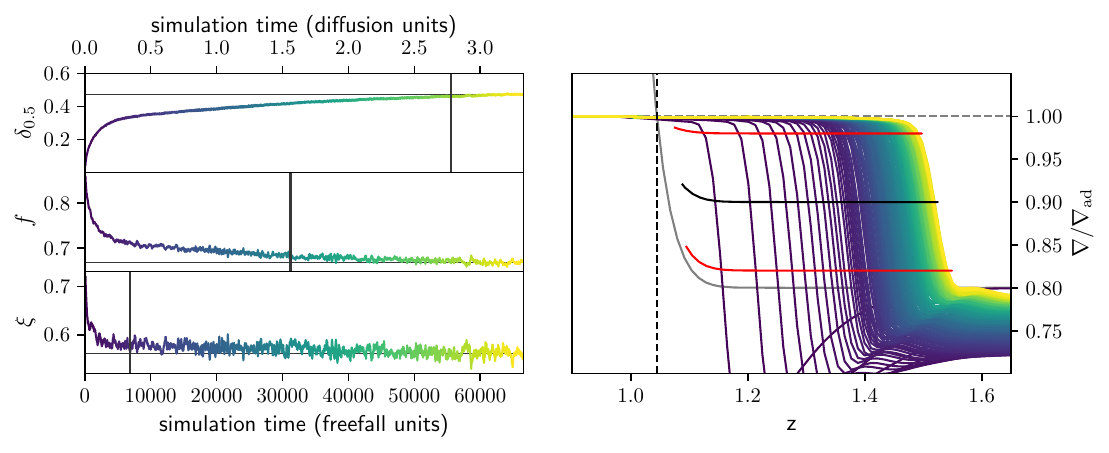}
\caption{
Time evolution of the landmark Case I simulation ($\mR = 400$, $\mP_D = 4$, $\mS = 10^3$).
In the top left panel, we plot the PZ height $\delta_{0.5}$ vs.~time.
Also shown are the time evolution of $f$ (middle left panel, defined in Eqn.~\ref{eqn:f_defn}) and $\xi$ (bottom left panel, defined in Eqn.~\ref{eqn:xi_defn}).
Thin horizontal lines denote the equilibrium values of each trace.
Vertical lines denote when each trace first converges to within 1\% of its equilibrium value.
(right panel) The vertical profile of $\justgrad/\gradad$ is plotted against height at regular time intervals.
The line color denotes the time, following the time traces in the left panels.
A horizontal dashed grey line denotes the constant value of $\gradad$.
The solid grey curve denotes the profile of $\gradrad$.
The location of the Schwarzschild convective boundary is displayed as a vertical dashed black line.
The top-of-PZ departure points (Eqn.~\ref{eqn:delta_p_measures}) are plotted over the profile evolution ($\delta_{0.1}$ and $\delta_{0.9}$ as red lines, $\delta_{0.5}$ as a black line).
\label{fig:time_evolution}
}
\end{figure*}

\section{Results}
\label{sec:results}

We now describe the results of the 3D dynamical simulations described in the previous section.
Fig.~\ref{fig:vertical_dynamics_panels} displays the dynamics in one of these simulations.
While we will briefly examine dynamics here, our primary goal in this section is to quantitatively compare our simulations to the theory of Sec.~\ref{sec:theory} using temporally averaged measures.

\subsection{Dynamics}
In Fig.~\ref{fig:lessturb_dynamics_panels} we display snapshots of the temperature anomalies in the two ``landmark'' simulations denoted by boxes in Fig.~\ref{fig:parameter_space}.
We display the temperature anomaly in the top panel of the Case I simulation with $\mR = 400$, $\mP_D = 4$, and $\mS = 10^3$; this simulation is included in all three of our parameter space sweeps and represents the point where our ($\mR, \mP, \mS$) cuts converge  in Fig.~\ref{fig:parameter_space}.
We display the temperature anomaly in the bottom panel of the Case II simulation with $\mR = 800$, $\mP_L = 4$, and $\mS = 10^3$.
The bulk Reynolds number in the convection zones of these simulations are (top) $\rm{Re} \sim 250$ and (bottom) $\rm{Re} \sim 350$.
Thus, these simulations are less turbulent than the simulation in Fig.~\ref{fig:vertical_dynamics_panels} (bulk Re $\sim$ 5000).
Aside from the degree of turbulence, the dynamics are very similar in Figs.~\ref{fig:vertical_dynamics_panels} \& \ref{fig:lessturb_dynamics_panels}.
In particular, we observe that \editone{relatively} hot plumes in the CZ turn into \editone{relatively} cold plumes in the PZ (as they cross the dashed horizontal lines), and \editone{relatively hot regions in the PZ lie above relatively cold regions in the CZ}.
Convective plumes extend through the penetrative region and impact the stable radiative zone (above the solid horizontal line).
The convective motions excite waves at a shallow angle above the stiff radiative-convective boundary.
We note that the Case II simulation has an additional temperature inversion at the base of the simulation.
Case II simulations have a linearly increasing conductivity $k$ in the convection zone, so there is formally a small penetrative region where $\justgrad \approx \gradad > \gradrad$ at the base of the domain below the internal heating layer (lower dotted line in bottom panel of Fig.~\ref{fig:lessturb_dynamics_panels}).

While the landmark simulations in Fig.~\ref{fig:lessturb_dynamics_panels} are not as turbulent as the dynamics in Fig.~\ref{fig:vertical_dynamics_panels}, they are sufficiently nonlinear to be interesting.
Importantly, these simulations develop large penetration zones, and can be evolved for tens of thousands of convective overturn times.
As we will demonstrate in the next section, the formation timescale of penetrative zones can take tens of thousands of convective overturn times.

\begin{figure*}[t]
\centering
\includegraphics[width=\textwidth]{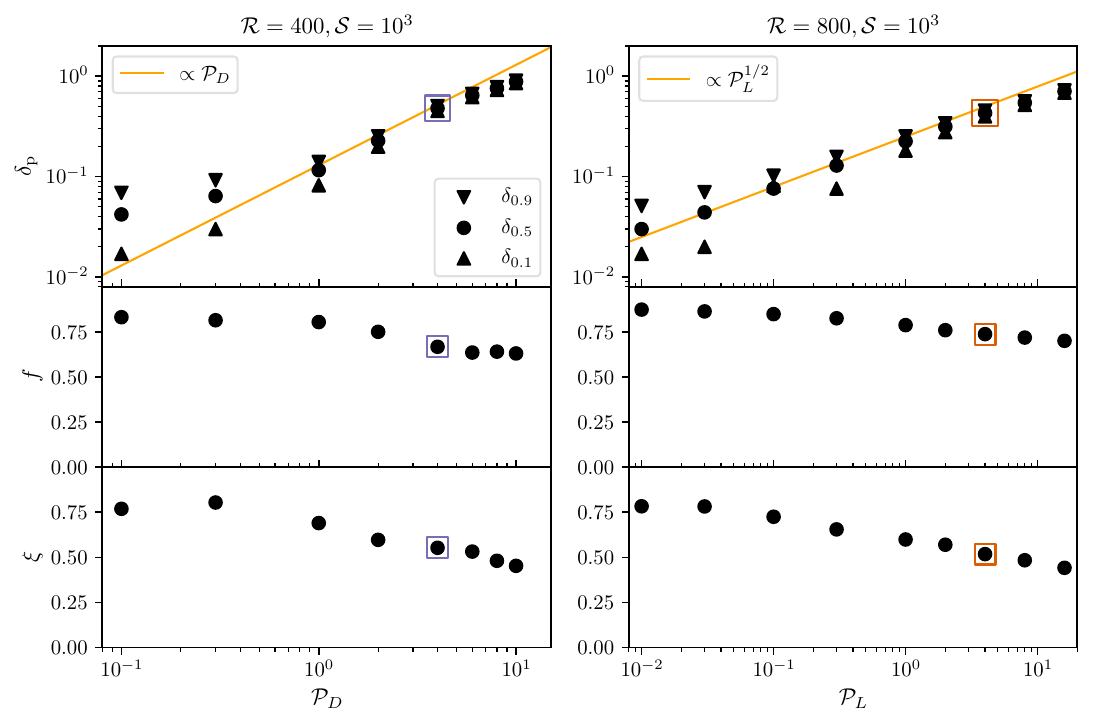}
\caption{
Simulation results vs.~$\mP$ for both Case I (left panels; solid purple circles in Fig.~\ref{fig:parameter_space}) and Case II (right panels; solid orange diamonds in Fig.~\ref{fig:parameter_space}).
Boxed data points denote landmark simulations from Fig.~\ref{fig:parameter_space}.
The top panels show the penetration height according to Eqn.~\ref{eqn:delta_p_measures}.
The Case I penetration heights (upper left) vary linearly with $\mP$, in line with the prediction of Eqn.~\ref{eqn:discontinuous_prediction}.
The Case II penetration heights (upper right) vary like $\sqrt{\mP}$, in line with the prediction of Eqn.~\ref{eqn:linear_prediction}.
In the middle panels, we measure $f$ according to Eqn.~\ref{eqn:f_defn}.
We find values of $f \in [0.6, 0.9]$, and changes in $f$ are secondary to changes in $\mP$ for determining penetration heights.
In the bottom panels, we measure $\xi$ according to Eqn.~\ref{eqn:xi_defn}.
We find characteristic values of $\xi \in [0.5, 0.75]$, suggesting that the falloff of the $\bar{\Phi}$ in the PZ is well described by a linear function (at high $\mP$ when $\xi \approx 1/2$), or by a cubic function (at low $\mP$ when $\xi \approx 3/4$).
\label{fig:parameters_vs_p}
}
\end{figure*}

\subsection{Qualitative description of simulation evolution}

In Fig.~\ref{fig:time_evolution}, we show the time evolution of the landmark Case I simulation ($\mR = 400$, $\mS = 10^3$, and $\mP_D = 4$) whose initial temperature profile sets $\justgrad = \gradad$ in the convection zone ($z \lesssim 1$) and $\justgrad = \gradrad$ in the radiative zone ($z \gtrsim 1$).
In the top left panel, we display the height of the penetrative region $\delta_{\rm{0.5}}$ vs.~time.
This region initially grows quickly over hundreds of freefall times, but this evolution slows down; reaching the final equilibrium takes tens of thousands of freefall times.
The evolution of the other parameters in our theory ($f$, $\xi$) are shown in the middle and bottom left panels of Fig.~\ref{fig:time_evolution}.
We plot the rolling mean, averaged over 200 freefall time units. 
We see that the values of $f$ and $\xi$ reach their final values ($f \approx 0.67$, $\xi \approx 0.58$) faster than the penetration zone evolves to its full height.
We quantify this fast evolution by plotting vertical lines in each of the left three panels corresponding to the first time at which the rolling average converges to within 1\% of its equilibrated value.
The equilibrated value is averaged over the final 1000 freefall times of the simulation and plotted as a grey horizontal line.
The evolved value of $f$ indicates that roughly 2/3 of the buoyancy driving is dissipated in the bulk CZ, so that 1/3 is available for PZ dissipation and negative buoyancy work.
The evolved value of $\xi$ indicates that the shape of dissipation in the PZ is slightly steeper than linear.

\begin{figure*}[t]
\centering
\includegraphics{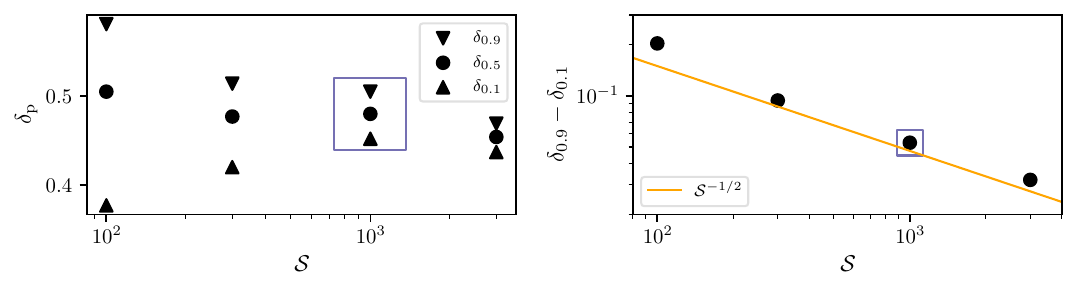}
\caption{
Case I simulations results vs.~$\mS$ at $\mR = 400$, $\mP = 4$.
Boxed data points denote the landmark simulation from Fig.~\ref{fig:parameter_space}.
(Left panel) Penetration heights vs.~$\mS$.
While $\delta_{0.1}$ and $\delta_{0.9}$ show some variation, the mean penetration height ($\delta_{0.5}$) is roughly constant.
(Right panel) The width of the thermal transition layer ($\delta_{0.9} - \delta_{0.1}$) vs.~$\mS$.
We roughly observe a $\mS^{-1/2}$ scaling.
\label{fig:parameters_vs_s}
}
\end{figure*}

In the right panel of Fig.~\ref{fig:time_evolution}, we plot the profile of $\justgrad/\gradad$ in our simulation at regular time intervals, where the color of the profile corresponds to time, as in the left panels.
$\gradad$ is plotted as a dashed horizontal line while $\gradrad$ is plotted as a grey solid line which decreases with height around $z \approx 1$ and satures to a constant above $z \gtrsim 1.1$.
The location of the Schwarzschild boundary, $L_s$, is overplotted as a black vertical dashed line.
We note that the Schwarzschild boundary does not move over the course of our simulation, so the extention of the convection zone past this point is true penetration and not the result of entrainment-induced changes in the Schwarzschild (or Ledoux) convective boundaries.
The traces of $\delta_{0.1}$ and $\delta_{0.9}$ are overplotted as red lines while that of $\delta_{0.5}$ is plotted as a black line.
We see that the fast initial evolution establishes a sizeable PZ (denoted by purple $\justgrad$ profiles), but its final equilibration takes much longer (indicated by the separation between the purple, green, and yellow profiles decreasing over time).

This long evolution is computationally expensive; for this modest simulation (256x64$^2$ coefficients), this evolution takes roughly 24 days on 1024 cores for a total of $\sim$600,000 cpu-hours.
It is not feasible to perform simulations of this length for a full parameter space study, and so we accelerate the evolution of most of the simulations in this work.
To do so, we take advantage of the nearly monotonic nature of the evolution of $\delp$ vs.~time displayed in Fig.~\ref{fig:time_evolution}.
We measure the instantaneous values of $(\delta_{0.1}, \delta_{0.5}, \delta_{0.9})$, as well as their instantaneous time derivatives.
Using these values, we take a large ``time step'' forward \editone{to evolve $\delp$.}
While doing so, we preserve the width of the transition from the PZ to the RZ, and we also adjust the solution so that $\justgrad = \gradrad$ in the RZ, effectively equilibrating the RZ instantaneously.
In other words, we reinitialize the simulation's temperature profile with a better guess at its evolved state based on its current dynamical evolution.
For details on how this procedure is carried out, see Appendix \ref{app:accelerated_evolution}.

\subsection{Dependence on $\mP$}
We find that the height of the penetration zone is strongly dependent on $\mP$.
In the upper two panels of Fig.~\ref{fig:parameters_vs_p}, we plot the penetration height ($\delta_{0.1}$, $\delta_{0.5}$, $\delta_{0.9}$ from Eq.~\ref{eqn:delta_p_measures}) from Case I simulations (discontinuous $k$, upper left) and Case II simulations (discontinuous $\partial_z k$, upper right).
The fixed values of $\mR$ and $\mS$ are shown above these panels.
We find that the leading-order $\mP$ scaling predictions of Eqns.~\ref{eqn:discontinuous_prediction} \& \ref{eqn:linear_prediction} describe the data \editone{well at intermediate values of $\mP$} (orange lines).
At small values of $\mP$ we see somewhat weaker scalings than these predictions, because the profiles of $k$ and $\partial_z k$ are not truly discontinuous but jump from one value in the CZ to another in the RZ over a finite width (see e.g., the $\gradrad$ profile in Figs.~\ref{fig:grad_profiles} \& \ref{fig:time_evolution} and Sec.~\ref{sct:numerics}).
At large values of $\mP$, the penetration height falls off of these predicted scaling laws.
In this regime, dissipation dominates over buoyancy in the PZ, so the PZ height saturates.

The middle and bottom panels of Fig.~\ref{fig:parameters_vs_p} demonstrate that that $f$ and $\xi$ are to leading order constant with $\mP$.
However, we find that $f$ has slightly smaller values in the Case I simulations (left) than in the Case II simulations (right).
We measure characteristic values of $f \in [0.6, 0.9]$, signifying that 60-90\% of the buoyant work is balanced by dissipation in the convection zone, depending on the simulation.
We note a weak trend where $f$ decreases as $\mP$ increases.
As $\mP$ increases, we find that CZ velocities decrease, leading to a decrease in the dissipation rate.
When $\mP$ is small, the PZ-RZ boundary (which acts like a wall, left panel of Fig.~\ref{fig:vertical_dynamics_panels}) efficiently deflects convective velocities sideways resulting in increased bulk-CZ velocities.
As $\mP$ grows, the velocities have access to an extended PZ in which to buoyantly \editone{decelerate} before deflection, resulting in slightly lower bulk velocities.
A similar trend of $\xi$ decreasing as $\mP$ increases can be seen.
Recall that smaller values of $\xi$ indicate the dissipative dynamics are rather different in the PZ and CZ.
As the size of the PZ grows, the dynamical structures of the PZ shift from what is found in the CZ, and so $\xi$ shrinks.

\subsection{Dependence on $\mS$}

We find that the height of the penetration zone is weakly dependent on $\mS$.
In the left panel of Fig.~\ref{fig:parameters_vs_s}, we plot the penetration height of a few Case I simulations with $\mP_D = 4$ and $\mR = 400$ but with different values of $\mS$.
The mean penetration height $\delta_{0.5}$ varies only weakly with changing $\mS$, but that the values of $\delta_{0.1}$ and $\delta_{0.9}$ vary more strongly.
The \editone{PZ-RZ boundary layer} in which $\justgrad$ changes from $\gradad$ to $\gradrad$ becomes narrower as $\mS$ increases.
To quantify this effect, we plot $\delta_{0.9} - \delta_{0.1}$ in the righthand panel of Fig.~\ref{fig:parameters_vs_s}.
We find that the width of this region varies roughly according to a $\mS^{-1/2}$ scaling law, reminiscent of the pure-overshoot law described by \citet{korre_etal_2019}.
%This suggests that the evolved profile of a convecting region can be described by a penetration zone (described by e.g., $\mP$, $f$, and $\xi$) with a thin overshooting region (described by $\mS$) within the transition between the PZ and RZ.

Note that if the enstrophy, $\omega^2$ in the convection zone exceeds the value of the square buoyancy frequency $N^2$ in the radiative zone, the gravity waves in the RZ become nonlinear.
We therefore restrict the simulations in this study to relatively large\footnote{These values are large for nonlinear simulations, but modest compared to astrophysical values. \editone{While there is observational uncertainty about the magnitude of deep convective velocities in the Sun, in the MESA model presented in Sct.~\ref{sec:solar_model}, $f_{\rm{conv}} \approx 10^{-6}$ s$^{-1}$ and $N \approx 10^{-3}$ s$^{-1}$, so $\mS \approx 10^6$.}} values of $10^{2} \leq \mS < 10^4$ in order to ensure $N^2 > \omega^2$ even in our highest enstrophy simulations.

\subsection{Dependence on $\mR$}

\begin{figure*}[t]
\centering
\includegraphics[width=\textwidth]{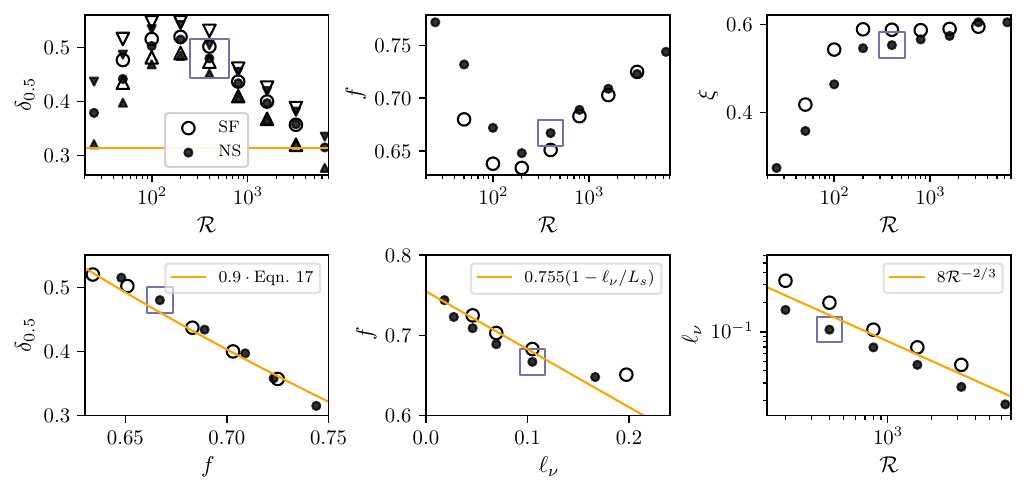}
\caption{
(Upper left panel) Penetration heights vs.~$\mR$ for Case I simulations (vertical cuts in Fig.~\ref{fig:parameter_space}).
Empty markers denote stress-free boundaries (SF) and filled markers denote no-slip boundaries (NS).
In both cases, we see a roughly logarithmic decrease of $\delp$ vs.~$\mR$.
(Upper middle panel) $f$ increases with $\mR$.
(Upper right panel) $\xi$ does not change appreciably with $\mR$ for turbulent simulations with $\mR \geq 200$.
(Lower left panel) There is a strong correlation between $\delta_{0.5}$ and $f$, agreeing with our theoretical model of Eqn.~\ref{eqn:discontinuous_prediction}.
(Lower middle panel) Changes in $f$ are roughly linearly proportional to the depth of the viscous boundary layer, $\ell_\nu$, at the bottom of the domain.
(Lower right panel) $\ell_\nu$ follows a well-known convective scaling law, so $\delta_{0.5}$ and $f$ should saturate as $\mR \rightarrow \infty$ and $\ell_{\nu} \rightarrow 0$.
Boxed data points denote the landmark simulation from Fig.~\ref{fig:parameter_space}.
\label{fig:parameters_vs_re}
}
\end{figure*}

We find that the height of the penetration zone is weakly dependent on $\mR$.
In the upper left panel of Fig.~\ref{fig:parameters_vs_re}, we find a logarithmic decrease in the penetration height with the Reynolds number.
In order to understand how this could happen at fixed $\mP$, we also plot the output values of $f$ (upper middle) and $\xi$ (upper right).
We find that $f$ increases with increasing $\mR$, but is perhaps leveling off as $\mR$ becomes large.
We find that $\xi$ does not increase strongly with $\mR$ except for in the case of laminar simulations with $\mR < 200$.
Eqn.~\ref{eqn:discontinuous_prediction} predicts that $\delp$ should change at fixed $\mP$ and $\xi$ if $f$ is changing.
In the bottom left panel, we show that the change in $\delp$ is due to this change in $f$.
We find that this is true both for simulations with stress-free dynamical boundary conditions (open symbols, SF) and for no-slip conditions (closed symbols, NS).

We now examine why $f$ increases as $\mR$ increases.
In the SF simulations, within the CZ, we can reasonably approximate $\bar{\Phi}$ as a constant $\Phi_{\rm{CZ}}$ in the bulk and zero within the viscous boundary layer,
\begin{equation}
\bar{\Phi}(z) = 
\begin{cases}
\Phi_{\rm{CZ}} 	& z > \ell_\nu \\
0				& z \leq \ell_\nu
\end{cases},
\label{eqn:cz_dissipation_model}
\end{equation}
where $\ell_\nu$ is the viscous boundary layer depth.
We have visualized a NS dissipation profile in the bottom panel of Fig.~\ref{fig:theory_profiles}; SF simulations look similar in the bulk, but drop towards zero at the bottom boundary rather than reaching a maximum.
Then, we have
\begin{equation}
\int_{\rm{CZ}} \bar{\Phi}\,dz \approx \Phi_{\rm{CZ}} \left( L_s - \ell_\nu \right),
\end{equation}
and so per Eqn.~\ref{eqn:f_defn},
\begin{equation}
f = f_\infty\left(1 - \frac{\ell_\nu}{L_s}\right),
\label{eqn:f_infty}
\end{equation}
where $f_\infty$ is the expected value of $f$ at $\mR = \infty$ when $\ell_\nu = 0$.
So we see that the CZ dissipation and therefore $f$ vary linearly with $\ell_\nu$.

In the bottom middle panel of Fig.~\ref{fig:parameters_vs_re}, we find that Eqn.~\ref{eqn:f_infty} with $f_\infty = 0.755$ captures the high-$\mR$ behavior.
To measure $\ell_\nu$, we first measure the height of the extremum of the viscous portion of the kinetic energy flux $\bar{\mathcal{F}}$ near the boundary, and take $\ell_\nu$ to be the twice that height.
We find that Eqn.~\ref{eqn:f_infty} is a slightly better description for the SF simulations than the NS simulations; NS simulations have maximized dissipation in the boundary layer, and therefore Eqn.~\ref{eqn:cz_dissipation_model} is a poor model for $z \leq \ell_\nu$.
In the bottom right panel of Fig.~\ref{fig:parameters_vs_re}, we demonstrate that the depth of the viscous boundary layer follows classical scaling laws from Rayleigh-B\'{e}nard convection\footnote{
If you assume the Nusselt Number dependence on the Rayleigh number is throttled by the boundaries, Nu $\propto$ Ra$^{1/3}$ (as is frequently measured), and the Reynolds number is Re $\propto$ Ra$^{1/2}$, you retrieve Nu $\propto$ Re$^{2/3}$. 
The Nusselt number generally varies like the inverse of the boundary layer depth, Nu $\propto$ $\ell^{-1}$, and so we expect $\ell_{\nu} \propto \mR^{-2/3}$.
} \citep{ahlers_etal_2009, goluskin2016}.
Combining these trends, we expect 
\begin{equation}
f = f_\infty(1 - C \mR^{-2/3})
\end{equation}
for a constant $C$.
Thus as $\mR \rightarrow \infty$, $f \rightarrow f_\infty$.

We use the fitted function of $f$ from the bottom middle panel, along with Eqn.~\ref{eqn:discontinuous_prediction}, to estimate $\delta_{0.5}$ in the bottom left panel.
We need to multiply this equation by a factor of 0.9, which accounts for some differences between the simulations and the idealized ``discontinuous flux'' theoretical model.
First, due to internal heating and the finite width of the conductivity transition around the Schwarzschild boundary, the convective flux is not truly constant through the full depth of the CZ.
Thus, we expect $L_{\rm{CZ}}$  in Eqn.~\ref{eqn:discontinuous_prediction} to be smaller than 1.
Furthermore, the theory is derived in the limit of an instantaneous transition from $\gradad$ to $\gradrad$ where $\delta_{0.1} = \delta_{0.5} = \delta_{0.9}$; our simulations have a finite transition width.
Despite these subtle differences, we find good agreement.

Using $f_\infty = 0.755$ we estimate that $\delta_{0.5} \approx 0.31$ for $\mR \rightarrow \infty$ and plot this as a horizontal orange line on the upper left panel of Fig.~\ref{fig:parameters_vs_re}.
This value is coincidentally very near the value of $\delta_{0.5}$ achieved in our highest-$\mR$ simulations.
Unfortunately, we cannot probe more turbulent simulations.
We can only run the $\mR = 6.4 \times 10^{3}$ simulation for a few hundred freefall times.
Our accuracy in measuring results from this simulation is limited by the long evolutionary timescales of the simulation (see Fig.~\ref{fig:time_evolution} for similar evolution in a less turbulent, $\mR = 400$ case).
Even accounting for our accelerated evolutionary procedure, we can only be confident that the PZ heights of this simulation are converged to within a few percent.
Future work should aim to better understand the trend of PZ height with turbulence.
However, the displayed relationships between $\delp$ and $f$, $f$ and $\ell_\nu$, and $\ell_\nu$ and $\mR$ --- all of which are effects we largely understand --- suggest that PZ heights should saturate at high $\mR$.

In summary, we find that $\delp$ decreases as $\mR$ increases.
We find that these changes are caused by increases in $f$.
In our simulations, $f$ seems to have a linear relationship with the size of the viscous boundary layer $\ell_\nu$.
By measuring $f$ and $\ell_\nu$ in a simulation, the value of $f_\infty$ can be found from Eqn.~\ref{eqn:f_infty}.
Stellar convection zones are not adjacent to hard walls\footnote{
Core convection zones have no lower boundary due to geometry; flows pass through the singular point at $r = 0$.
Convective shells in should be bounded both above and below by penetrative regions.
}, so $f_\infty$ and the limit $\ell_\nu \rightarrow 0$ applies to stellar convection.

While we have examined a Case I simulation with $\mP = 4$ here, we expect the simulation with $\mP_L = 1$ (a linear radiative conductivity profile) to be the most representative of conditions near a stellar convective boundary.
In this simulation, we measure $\xi \approx 0.6$, $f \approx 0.785$, $\ell_\nu \approx 0.08$, and $L_s = 1$.
Using Eqn.~\ref{eqn:f_infty}, we estimate that
\begin{equation}
f_\infty = 0.86\qquad\rm{and}\qquad
\xi = 0.6
\label{eqn:f_xi_estimates}
\end{equation}
are good first estimates for $f$ and $\xi$ when applying our theory of penetrative convection to stellar models.

\section{Testing our parameterization in a simple stellar model of the Sun}
\label{sec:solar_model}

\begin{figure}[t]
\centering
\includegraphics[width=\columnwidth]{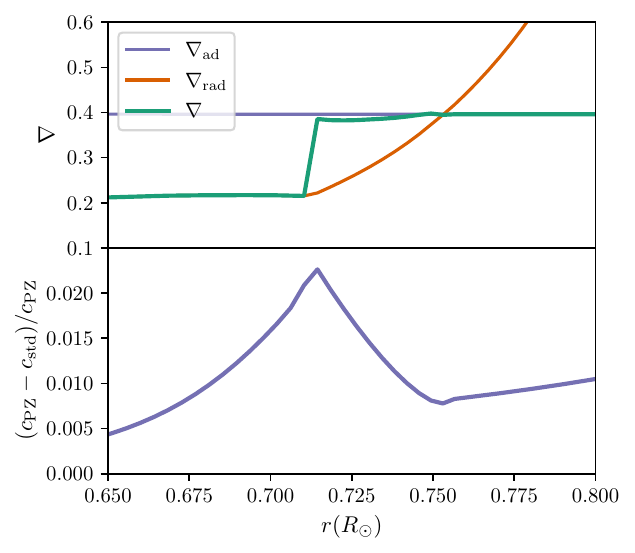}
\caption{
(top) Profiles of $\justgrad$ (green), $\gradad$ (purple), and $\gradrad$ (orange) in a 1 $M_\odot$ MESA stellar model with a penetration zone.
(bottom) Sound speed differences between the model shown in the top panel and a standard (std) model run at identical parameters but without a PZ.
The addition of a PZ creates an acoustic glitch, raising the sound speed by $\mathcal{O}$(2\%) below the convection zone.
\label{fig:mesa_profiles}
}
\end{figure}

Our simulation results present a strong case for a flux- and dissipation-based model of convective penetration, similar to those considered by \citet{zahn1991} and \citet{roxburgh1989}.
In this section, we discuss a simple stellar model of the Sun which we have created by implementing our parameterization into MESA (see Appendix~\ref{app:mesa}).
We of course note that the theory and 3D simulations in this work do not include many of the complications of stellar convection like density stratification, sphericity, rotation, magnetism, etc.
We present this model as a proof of concept and to inspire further work.

In order to implement our theory into MESA, we need to extend Eqn.~\ref{eqn:theory_fraction} to spherical geometry.
To do so, we replace horizontal averages in Eqn.~\ref{eqn:integral_constraint} with integrals over latitude and longitude, and find that the relevant integral constraint contains the convective luminosity,
\begin{equation}
\int |\alpha| g L_{\rm{conv}}\,dr =   \int_V \rho_0 \Phi\,dV,
\label{eqn:mesa_integral_constraint}
\end{equation}
where $L_{\rm{conv}} = 4\pi\rho_0 r^2 \bar{F_{\rm{conv}}}$, $r$ is the radial coordinate, and we write the RHS as a volume integral.
We next define $f$ in the same way as in Eqn.~\ref{eqn:f_defn} and define $\xi$ similarly to Eqn.~\ref{eqn:xi_defn},
\begin{equation}
\int_{\rm{PZ}} \rho_0 \Phi \,dV = \xi \frac{V_{\rm{PZ}}}{V_{\rm{CZ}}}\int_{\rm{CZ}}\rho_0 \Phi \, dV,
\label{eqn:sphere_xi_defn}
\end{equation}
where $V_{\rm{PZ}}$ and $V_{\rm{CZ}}$ are the volumes of the PZ and CZ respectively.
Eqn.~\ref{eqn:sphere_xi_defn} generalizes Eqn.~\ref{eqn:xi_defn} outside of the assumption of a plane-parallel atmosphere.
Thus Eqn.~\ref{eqn:theory_fraction} in spherical geometry is
\begin{equation}
-\frac{\int_{\rm{PZ}} L_{\rm{conv}}\,dr}{\int_{\rm{CZ}} L_{\rm{conv}}\,dr} + f \xi \frac{V_{\rm{PZ}}}{V_{\rm{CZ}}} = (1 - f),
\label{eqn:mesa_eqn}
\end{equation}

We implemented Eqn.~\ref{eqn:mesa_eqn} in MESA (see Appendix \ref{app:mesa} for details) and evolved a $1 M_\odot$ model to an age of 4.56 Gyr with $f = 0.86$ and $\xi = 0.6$ (Eqn.~\ref{eqn:f_xi_estimates}) to qualitatively understand how our penetration parameterization modifies a stellar model.
In the top panel of Fig.~\ref{fig:mesa_profiles} we display $\justgrad \equiv d\ln T/d\ln P$ from the model which includes convective penetration.
Note that $\justgrad$ (green) remains close to $\gradad$ (purple) below the Schwarzschild convective boundary ($\gradad = \gradrad$) in a penetration zone.
After some depth $\justgrad \rightarrow \gradrad$ (orange) in the star's interior.
We additionally evolved a standard 1 $M_\odot$ MESA model to a 4.56 Gyr age without the inclusion of a PZ.
We compare the sound speed $c$ profiles of the PZ and standard (std) model in the bottom panel of Fig.~\ref{fig:mesa_profiles}.
When a PZ is present beneath a CZ, $\justgrad$ experiences a sharp jump from $\gradad$ to $\gradrad$ (Fig.~\ref{fig:mesa_profiles}, top panel), resulting in an acoustic ``glitch'' in the sound speed profile.

In the model shown in Fig.~\ref{fig:mesa_profiles}, we find $H_p \approx 0.082R_\odot$ at the Schwarzschild CZ boundary, and the depth of the penetration zone in Fig.~\ref{fig:mesa_profiles} is $0.042R_\odot \sim 0.5 H_p$.
The inclusion of this PZ leads to an $\mathcal{O}$(2\%) increase in $c$ near the base of the solar convection zone.
Helioseismic observations suggest a similar increase below the base of the solar convection zone \citep[e.g.,][their Fig. 17]{christensen-dalsgaard_etal_2011}.
The difference $\Delta c = c_{\rm{PZ}} - c_{\rm{std}}$ that we see in this stellar model of the Sun (Fig.~\ref{fig:mesa_profiles}) has the same sign and roughly the same shape.
However, the magnitude of the change in $c$ is larger than is observed; literature values include $\Delta c / c \approx \mathcal{O}(1\%)$ \citep{bergemann_serenelli_2014} and $\Delta c^2 / c^2 \approx \mathcal{O}(0.4\%)$ \citep{christensen-dalsgaard_etal_2011}, and our sound speed bump is located at a different radius than the observed bump.
Other helioseismic studies have argued that that the solar PZ depth cannot be larger than $\mathcal{O}$(0.05 $H_p$), because larger PZs would result in larger glitches than are detected  \citep[see Sct.~7.2.1 of][for a nice review]{basu2016}.
It is interesting, however, that the width of the PZ in Fig.~\ref{fig:mesa_profiles} is strikingly similar to the inferred width of the tachocline $(0.039 \pm 0.013)R_\odot$ that is reported by \citet{charbonneau_etal_1999}.

It is unsurprising that our Boussinesq-based model only qualitatively matches observational constraints for the solar CZ.
The solar convection zone is highly stratified ($\sim$14 density scale heights), and we neglected density stratification in this work.
Furthermore, the solar model used here is essentially a ``stock'' MESA model and has obvious disagreements with the solar model~S \citep[see Fig.~1 in][where the Schwarzschild base of the CZ is $r/R_\odot \approx 0.712$, whereas the one in Fig.~\ref{fig:mesa_profiles} is at $r/R_{\odot} \approx 0.75$]{christensen-dalsgaard_etal_2011}.
Despite the limitations of this minimal proof of concept, Fig.~\ref{fig:mesa_profiles} shows that our parameterization can produce penetration zones in 1D models with measurable acoustic glitches.
In a future paper, we will produce more realistic models by building upon our parameterization to include the crucial effects of density stratification.
\editone{
    We note briefly that the theory in e.g., Eqn.~\ref{eqn:mesa_eqn} only knows about integral quantities of the convection and does not therefore know about quantities like the filling factor of upflows and downflows which stratification would modify.
    We suspect that dynamical differences that arise from including stratification would manifest as changes in $f$ and $\xi$, but a detailed exploration is beyond the scope of this work.
}

\section{Discussion}
\label{sec:discussion}
In this work, we presented dynamical simulations of convective penetration, in which convection mixes ${\justgrad \rightarrow \gradad}$ beyond the Schwarzschild boundary.
To understand these simulations, we used an integral constraint \citep[reminiscent of][]{roxburgh1989} and flux-based arguments \citep[similar to][]{zahn1991} to derive a parameterization of convective penetration according to the convective flux and viscous dissipation.
In doing so, we have laid down the first steps (Eqns.~\ref{eqn:theory_fraction} \& \ref{eqn:mesa_eqn}) towards incorporating convective penetration into stellar structure codes.
We parameterized the viscous dissipation into a bulk-CZ portion ($f$) and a portion in the extended penetrative region ($\xi$), and derived predictions for how the height of a penetrative region $\delp$ should scale with these measurable parameters and a new flux-based ``penetration parameter'' $\mP$.
We designed and analyzed two sets of simulations which showed good agreement with these theoretical predictions.
\editone{
    These simulations differ from past studies because we separately specify $\mP$ and the stiffness $\mS$, and we allow the simulations to evolve for a very long time or use numerical techniques for rapid evolution.
}
We briefly examined what the impliciations of this theory could be for a simple stellar model.

Our simulation results suggest that stellar convection zones could be bounded by sizeable penetration zones.
In extreme simulations, we observe penetration zones which are as large as the convection zones they accompany; however, for realistic stellar values ($\mP \approx 1$), we find that they may be as large as 20-30\% of the convective zone length scale ($\sim$the mixing length).

The simulations we presented in this work use a simplified setup to test the basic tenets of our theory.
In particular, they demonstrate that the shape of the flux near the convective boundary and the viscous dissipation together determine the height of the penetration zone.
The precise values of the parameters $f$ and $\xi$ achieved in natural, turbulent, fully compressible, spherical stellar convection may be different from those presented in e.g., Fig.~\ref{fig:parameters_vs_p} and Eqn.~\ref{eqn:f_xi_estimates} here.
Future work should aim to understand how these parameters and the theory presented in e.g., Eqn.~\ref{eqn:mesa_eqn} change when more realistic effects are taken into account.

Stellar opacities and thus stellar radiative conductivities are functions of thermodynamic variables rather than radial location.
The formation of a penetration zone will therefore affect the conductivity profile and $\gradrad$, which will in turn affect the location of the Schwarzschild boundary and the estimate of how deep the penetration zone should be.
In other words, convective penetration and entrainment \emph{both} occur in realistic settings, and their combined effects should be studied.
Future work should follow e.g., \citet{kapyla_etal_2017} and implement realistic opacity profiles which evolve self-consistently with the thermodynamic state in order to understand how these effects feedback into one another.

\edittwo{
    Our simulation setup (in which convection is driven by internal heating and stopped by a radiative flux divergence) most closely imitates core convection in massive stars.
    Other shell or envelope convection zones in stars are driven entirely by divergences in the radiative flux.
    These divergences act as radiative heating (at the base of the convection zone) and radiative cooling (at the top of the convection zone).
    We suspect that our simulation setup (and separate specification of $\mP$ and $\mS$) could straightforwardly be implemented in a model where the total flux is constant with height and convection is driven entirely by changes in $k$ with height.
    Future work should test this by examining three-layer experiments where a CZ sits between two RZs, and the convection is driven at the base by a decrease in $k$ and then stopped by an increase in $k$ at the top.
    These experiments would help constrain how penetration zone depths change when \emph{two} PZs (one above, one below) must be accounted for in the integral constraint.
}

Our work here assumes a uniform composition through the convective and radiative region.
Convective boundaries often coincide with discontinuities in composition profiles \citep{salaris_cassisi_2017}.
Future work should determine if stabilizing composition gradients can prevent the formation of the penetration zones seen here.

Furthermore, stellar fluid dynamics exist in the regime of Pr$\,\ll1$ \citep{garaud2021}.
Dynamics in this regime may be different from those in the regime of Pr $\lesssim 1$ that we studied here, which in theory could affect $f$ and $\xi$.
Recently, \citet{kapyla2021} found that convective flows exhibited more penetration at low Pr than high Pr.
Future work should aim to understand whether $f$ and/or $\xi$ depend strongly on $\Pran$ in the turbulent regime.

Two other interesting complications in stellar contexts are rotation and magnetism.
In the rapidly rotating limit, rotation creates quasi-two-dimensional flows, which could affect the length scales on which dissipation acts and thus modify $f$.
Furthermore, magnetism adds an additional ohmic dissipation term, which could in theory drastically change our hydrodynamical measurement of $f$.

In summary, we have unified \citet{roxburgh1989}'s integral constraint with \citet{zahn1991}'s theory of flux-dependent penetration into a parameterized theory of convective penetration.
We tested this theory with simulations and found good agreement between the theory and our simulations.
In future work, we will use simulations to test some of the complicating factors we discussed here and aim to more robustly implement convective penetration into MESA.

\begin{acknowledgments}
We thank Keaton Burns, Matt Browning, Matteo Cantiello, Geoff Vasil, and Kyle Augustson for useful discussions and/or questions which improved the content of this manuscript.
Ben Brown thanks Jeffrey Oishi for many years of discussions about overshooting convection.
    \edittwo{
        We thank the anonymous referee for carefully reading our manuscript, engaging with our science, and helping identify places where our descriptions of our simulations were confusing.
}EHA is funded as a CIERA Postdoctoral fellow and would like to thank CIERA and Northwestern University. 
We acknowledge the hospitality of Nordita during the program ``The Shifting Paradigm of Stellar Convection: From Mixing Length Concepts to Realistic Turbulence Modelling," where the groundwork for this paper was set.
This work was supported by NASA HTMS grant 80NSSC20K1280 and NASA SSW grant 80NSSC19K0026.
Computations were conducted with support from the NASA High End Computing (HEC) Program through the NASA Advanced Supercomputing (NAS) Division at Ames Research Center on Pleiades with allocation GID s2276.
The Flatiron Institute is supported by the Simons Foundation.
\end{acknowledgments}

\appendix

\section{Accelerated Evolution}
\label{app:accelerated_evolution}
As demonstrated in Fig.~\ref{fig:time_evolution}, the time evolution of simulations which start from a state based on the Schwarzschild criterion can be prohibitively long.
In \citet{anders_etal_2018}, we explored the long time evolution of simple convective simulations and found that fast-forwarding the evolution of a convective simulation's internal energy and thermal structure can be done accurately.
This can be done because the convective dynamics converge rapidly even if the thermal profile converges slowly.
This same separation of scales is observed in the penetrative dynamics in this work, and so similar techniques should be applicable.

To more quickly determine the final size of the evolved penetration zones we use the following algorithm.
\begin{enumerate}
\item Once a simulation has a volume-averaged Reynolds number greater than 1, we wait 10 freefall times to allow dynamical transients to pass.
\item We measure the departure points ($\delta_{0.1}$, $\delta_{0.5}$, $\delta_{0.9}$) every freefall time, and store this information for 30 freefall times.
\item We linearly fit each of the departure points' evolution against time using NumPy's \texttt{polyfit} function.
We assume that convective motions influence $\delta_{0.1}$ and $\delta_{0.5}$ more strongly than $\delta_{0.9}$.
We measure the time-evolution of the convective front $\frac{d \delp}{dt}$ by averaging the slope of the linear fits for $\delta_{0.1}$ and $\delta_{0.5}$.
\item We take a large ``time step'' of size $\tau_{\rm{AE}}$ forward.
We calculate $\Delta \delta_p = \tau_{\rm{AE}} \frac{d \delp}{dt}$.
\begin{itemize}
\item If $\Delta \delta_p < 0.005$, we erase the first 15 time units worth of departure point measures and return to step 2 for 15 time units.
\item  If $\Delta \delta_p$ is large, we adjust the top of the PZ by setting $\delta_{0.5,\rm{new}} = \angles{\delta_{0.5}}_t + \Delta \delta_p$ (angles represent a time average).
If $|\Delta \delta_p| > 0.05$, we limit its value to 0.05.
We calculate the width of the \editone{PZ-RZ boundary layer} $d_w$ as the minimum of $\angles{\delta_{0.9} - \delta_{0.5}}_t$ and $\angles{\delta_{0.5} - \delta_{0.1}}_t$.
We adjust the mean temperature gradient to
\begin{equation}
\justgrad = \gradad + H(z; \delta_{\rm{0.5,\rm{new}}}, d_w) \Delta\justgrad,
\label{eqn:initial_grad}
\end{equation}
where $H$ is defined in Eqn.~\ref{eqn:heaviside} and $\Delta\justgrad = \gradrad - \gradad$.
We also multiply the temperature perturbations and full convective velocity field by $(1 - H(z; 1, 0.05))$.
This sets all fluctuations above the nominal Schwarzschild convection zone to zero, thereby avoiding any strange dynamical transients caused by the old dynamics at the radiative-convective boundary (which has moved as a result of this process).
\end{itemize}
\item Return to step 1.
\end{enumerate}
In general, the initial profile of $\bar{T}$ that we use when we start our simulations is given by Eqn.~\ref{eqn:initial_grad} with a value $\delta_{0.5,\rm{new}} \geq 0$.
We then evolve $\bar{T}$ towards a statistically stationary state using the above algorithm and standard timestepping.
If a simulation returns to step 2 from step 4 ten times over the course of its evolution, we assume that it has converged near its answer, stop this iterative loop, and allow the simulation to timestep normally.
Additionally, in some simulations, we ensure that this process occurs no more than 25 times.
This process effectively removes the long diffusive thermal evolution on display in the upper left panel of Fig.~\ref{fig:time_evolution} by immediately setting the mean temperature profile to the radiative profile above the PZ.

\begin{figure*}[t]
\centering
\includegraphics[width=\textwidth]{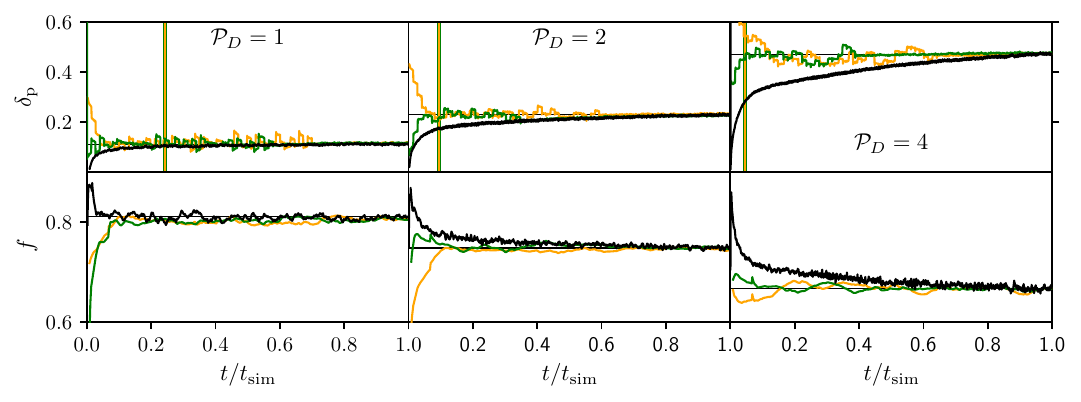}
\caption{
\label{fig:AE_time_figure}
(top row) Time traces of $\delta_{0.5}$ for simulations using standard timestepping (black lines), accelerated simulations with large initial values of $\delp$ (orange lines), and accelerated simulations with small initial values of $\delp$ (green lines).
Thin horizontal lines denote the equilibrated value of $\delta_{0.5}$.
Accelerated evolution timesteps can be seen as jumps in the $\delp$ trace.
After converging to within a few percent, the accelerated evolution procedure ``jitters'' around the equilibrated value.
Time units are normalized by the total run time of the simulation.
Accelerated simulations were run for $t_{\rm{sim}} = 3000$ freefall times.
The standard timestepping (black line) simulations were run for $t_{\rm{sim}} = 1.2 \times 10^4$ ($\mP_D = 1$), $t_{\rm{sim}} = 3.2 \times 10^4$ ($\mP_D = 2$), and $t_{\rm{sim}} = 6.7 \times 10^4$ ($\mP_D = 4$) freefall times.
The vertical green-and-yellow lines show the total simulation time of the accelerated simulation in terms of the direct simulation time; i.e., the accelerated simulation converged in only $\sim$ 5\% of the simulation time of the direct simulation for $\mP = 4$.
(Bottom row) Rolling average of $f$ over 200 freefall times, plotted in the same way as $\delta_{0.5}$.
}
\end{figure*}

In Fig.~\ref{fig:AE_time_figure}, we plot in black the time evolution of $\delp$ and $f$ in Case I simulations with $\mS = 10^3$, $\mR = 400$, and $\mP_D = [1,2,4]$.
We overplot the evolution of simulations which use this accelerated evolution (AE) procedure using orange and green lines.
Time units on the x-axis are normalized in terms of the total simulation run time in order to more thoroughly demonstrate the evolutionary differences between standard timestepping and AE.
However, the AE simulations are much shorter: the vertical green-and-yellow lines demonstrate how long the AE simulation ran compared to the standard timestepping simulation (so for $\mP_D = 1$, the AE simulations only took $\sim 1/4$ as long; for $\mP_D = 2$, they took $\sim 1/10$ as long; for $\mP_D = 4$, they took $\sim 1/20$ as long).
AE simulations with orange lines start with PZ heights which are much larger than the final height, while green line solution start with initial PZ heights which are smaller than the expected height.
Regardless of our choice of initial condition, we find that this AE procedure quickly evolves our simulations to within a few percent of the final value.
After converging to within a few percent of the proper penetration zone height, this AE procedure continues to iteratively ``jitter'' around the right answer until the convergence criterion we described above are met.
These jitters can be seen in the top panels of Fig.~\ref{fig:AE_time_figure}, where the solution jumps away from the proper answer in one AE iteration before jumping back towards it in the next iteration.
If the PZ height continues to noticeably vary on timescales of a few hundred freefall times, we continue to timestep the simulations until the changes of $\delp$ have diminished.

\section{MESA implementation}
\label{app:mesa}

Our 1D stellar evolution calculations were performed using the Modules for Experiments in Stellar Astrophysics software instrument \citep[MESA]{paxton_etal_2011, paxton_etal_2013, paxton_etal_2015, paxton_etal_2018, paxton_etal_2019}.

\subsection{Input Physics}

The MESA EOS is a blend of the OPAL \citep{Rogers2002}, SCVH
\citep{Saumon1995}, FreeEOS \citep{Irwin2004}, HELM \citep{Timmes2000},
and PC \citep{Potekhin2010} EOSes.

Radiative opacities are primarily from OPAL \citep{Iglesias1993,
Iglesias1996}, with low-temperature data from \citet{Ferguson2005}
and the high-temperature, Compton-scattering dominated regime by
\citet{Buchler1976}.  Electron conduction opacities are from
\citet{Cassisi2007}.

Nuclear reaction rates are from JINA REACLIB \citep{Cyburt2010} plus
additional tabulated weak reaction rates \citet{Fuller1985, Oda1994,
Langanke2000}.  (For MESA versions before 11701): Screening is
included via the prescriptions of \citet{Salpeter1954, Dewitt1973,
Alastuey1978, Itoh1979}.

\subsection{Penetration Implementation}

Here we describe a first implementation of Eqn.~\ref{eqn:mesa_eqn} in MESA.
We note that this impelementation is likely not universal or robust enough to be used in most complex stellar models, but it is robust enough to time-step stably and produce the results displayed in Sct.~\ref{sec:solar_model}.
Future work should improve upon this model.

To find the extent of the penetrative region we write Eqn.~\eqref{eqn:mesa_eqn} as
\begin{align}
	(1-f) \int_{\rm CZ} L_{\rm conv} dr = \int_{\rm PZ} \left(\xi f L_{\rm conv,avg,CZ} + L_{\rm conv}\right) dr,
	\label{eq:rwt}
\end{align}
where $L_{\rm conv,avg,CZ}$ is the average of $L_{\rm conv}$ in the convection zone and $L_{\rm conv}$ in the penetrative region is given by
\begin{align}
	L_{\rm conv} = \frac{L_{\rm rad}}{\nabla_r}(\nabla_a - \nabla_r),
\end{align}
which is the excess luminosity carried if the temperature gradient in the radiative zone is adiabatic.

We first integrate the left-hand side of Eqn.~\eqref{eq:rwt} over the convection zone and further use that to evaluate $L_{\rm conv,avg,CZ}$.
Next we integrate the right-hand side of the same away from the convective boundary into the radiative zone until the equation is satisfied.
The point where this integration stops is the edge of the penetrative region.

We then implement convective penetration in stellar evolution with two modifications.
First, we add an extra chemical mixing term in the penetration zone with a scale of $D \approx H_p (L/4\pi r^2 \rho)^{1/3}$, which is roughly the scale of the convective diffusivity.
The precise choice of diffusivity here does not matter, as any plausible scale will be enough to eliminate any composition gradient on evolutionary time-scales.
Secondly, we override the default routine in MESA for determining $\nabla$ and instead have the solver reduce $\nabla_a-\nabla$ by 90 per~cent in the penetrative zone.

Using this procedure with $f = 0.86$ and $\xi = 0.6$, and timestepping a solar model to the age of the current Sun ($\sim$ 4.5 Gyr), we find the profile displayed in Sec.~\ref{sec:solar_model}.

\subsection{Models}

Models were constructed to reasonably reproduce the present-day Sun and based on the 2019 MESA summer school lab by~\citet{pm}.
Inlists and the \texttt{run\_star\_extras} source code are available in a Zenodo repository \citep{supp}.

\section{Table of simulation parameters}
\label{app:simulation_table}
Input parameters and summary statistics of the simulations presented in this work are shown in Table~\ref{table:simulation_info}.

\begin{deluxetable*}{c c c c c c c c c c}
\tabletypesize{\footnotesize}
\caption{Table of simulation information.
\label{table:simulation_info}
}
\tablehead{
\colhead{Type} 	& \colhead{$\mP$}	& \colhead{$\mS$}	& \colhead{$\mR$}	& \colhead{$nx \times ny \times nz$}	&	\colhead{$t_{\rm{sim}}$}	&	\colhead{$(\delta_{0.1}, \delta_{0.5}, \delta_{0.9})$}	& \colhead{$f$}	& \colhead{$\xi$}	& \colhead{$\angles{u}$}
}
\startdata
\multicolumn{3}{l}{\textbf{``Standard timestepping'' simulations}}\\
D & $1.0$ & $ 10^{3}$ & $ 4.0 \cdot 10^{2}$ &  64x64x256   & $ 12347$ & (0.078, 0.112, 0.136) &  0.810 &  0.682 &  0.618 \\
D & $ 2.0$ & $ 10^{3}$ & $ 4.0 \cdot 10^{2}$ &  64x64x256   & $ 32057$ & (0.200, 0.230, 0.254) &  0.749 &  0.601 &  0.639 \\
D & $ 4.0$ & $ 10^{3}$ & $ 4.0 \cdot 10^{2}$ &  64x64x256   & $ 66557$ & (0.445, 0.472, 0.496) &  0.668 &  0.562 &  0.619 \\
\hline
\multicolumn{3}{l}{\textbf{``Accelerated Evolution'' simulations}}\\
D & $ 4.0$ & $ 10^{2}$ & $ 4.0 \cdot 10^{2}$ &  64x64x256   & $ 5000 $ & (0.377, 0.505, 0.581) &  0.654 &  0.526 &  0.617 \\
D & $ 4.0$ & $3.0 \cdot 10^{2}$ & $ 4.0 \cdot 10^{2}$ &  64x64x256   & $ 5000 $ & (0.420, 0.477, 0.514) &  0.663 &  0.551 &  0.618 \\
D & $ 10^{-1}$ & $ 10^{3}$ & $ 4.0 \cdot 10^{2}$ &  64x64x256   & $ 4561 $ & (0.017, 0.042, 0.069) &  0.831 &  0.769 &  0.588 \\
D & $ 3.0 \cdot 10^{-1}$ & $ 10^{3}$ & $ 4.0 \cdot 10^{2}$ &  64x64x256   & $ 4681 $ & (0.030, 0.064, 0.092) &  0.814 &  0.804 &  0.620 \\
D & $1.0$ & $ 10^{3}$ & $ 4.0 \cdot 10^{2}$ &  64x64x256   & $ 3000 $ & (0.082, 0.116, 0.140) &  0.804 &  0.690 &  0.624 \\
D & $ 2.0$ & $ 10^{3}$ & $ 4.0 \cdot 10^{2}$ &  64x64x256   & $ 5000 $ & (0.199, 0.228, 0.252) &  0.750 &  0.597 &  0.638 \\
D & $ 4.0$ & $ 10^{3}$ & $ 2.5 \cdot 10^{1}$ &  16x16x256   & $ 3000 $ & (0.321, 0.379, 0.437) &  0.772 &  0.274 &  0.343 \\
D & $ 4.0$ & $ 10^{3}$ & $ 5.0 \cdot 10^{1}$ &  32x32x256   & $ 3000 $ & (0.398, 0.442, 0.487) &  0.732 &  0.358 &  0.423 \\
D & $ 4.0$ & $ 10^{3}$ & $ 10^{2}$ &  32x32x256   & $ 3000 $ & (0.469, 0.503, 0.534) &  0.672 &  0.464 &  0.484 \\
D & $ 4.0$ & $ 10^{3}$ & $ 2.0 \cdot 10^{2}$ &  64x64x256   & $ 3000 $ & (0.485, 0.515, 0.542) &  0.648 &  0.546 &  0.548 \\
D & $ 4.0$ & $ 10^{3}$ & $ 4.0 \cdot 10^{2}$ &  64x64x256   & $ 5000 $ & (0.452, 0.480, 0.505) &  0.667 &  0.553 &  0.617 \\
D & $ 4.0$ & $ 10^{3}$ & $ 8.0 \cdot 10^{2}$ &  128x128x256 & $ 3000 $ & (0.407, 0.434, 0.455) &  0.689 &  0.566 &  0.678 \\
D & $ 4.0$ & $ 10^{3}$ & $ 1.6 \cdot 10^{3}$ &  128x128x256 & $ 3000 $ & (0.366, 0.397, 0.419) &  0.709 &  0.574 &  0.720 \\
D & $ 4.0$ & $ 10^{3}$ & $ 3.2 \cdot 10^{3}$ &  256x256x256 & $ 3235 $ & (0.321, 0.358, 0.381) &  0.723 &  0.605 &  0.746 \\
D & $ 4.0$ & $ 10^{3}$ & $ 6.4 \cdot 10^{3}$ &  384x384x384 & $ 414  $ & (0.277, 0.315, 0.335) &  0.744 &  0.605 &  0.757 \\
D & $ 6.0$ & $ 10^{3}$ & $ 4.0 \cdot 10^{2}$ &  64x64x256   & $ 6000 $ & (0.620, 0.647, 0.667) &  0.635 &  0.532 &  0.597 \\
D & $ 8.0$ & $ 10^{3}$ & $ 4.0 \cdot 10^{2}$ &  128x128x512 & $ 4357 $ & (0.732, 0.759, 0.779) &  0.640 &  0.481 &  0.592 \\
D & $ 10^{1}$ & $ 10^{3}$ & $ 4.0 \cdot 10^{2}$ &  128x128x512 & $ 4226 $ & (0.858, 0.885, 0.904) &  0.630 &  0.453 &  0.587 \\
D & $ 4.0$ & $3.0 \cdot 10^{3}$ & $ 4.0 \cdot 10^{2}$ &  64x64x512   & $ 1170 $ & (0.437, 0.454, 0.469) &  0.672 &  0.581 &  0.619 \\
D/SF & $ 4.0$ & $ 10^{3}$ & $ 5.0 \cdot 10^{1}$ &  32x32x256   & $ 5000 $ & (0.435, 0.477, 0.516) &  0.680 &  0.418 &  0.505 \\
D/SF & $ 4.0$ & $ 10^{3}$ & $ 10^{2}$ &  32x32x256   & $ 5000 $ & (0.482, 0.516, 0.547) &  0.638 &  0.543 &  0.573 \\
D/SF & $ 4.0$ & $ 10^{3}$ & $ 2.0 \cdot 10^{2}$ &  64x64x256   & $ 5000 $ & (0.490, 0.520, 0.547) &  0.634 &  0.589 &  0.640 \\
D/SF & $ 4.0$ & $ 10^{3}$ & $ 4.0 \cdot 10^{2}$ &  64x64x256   & $ 8000 $ & (0.474, 0.502, 0.531) &  0.651 &  0.588 &  0.693 \\
D/SF & $ 4.0$ & $ 10^{3}$ & $ 8.0 \cdot 10^{2}$ &  128x128x256 & $ 5000 $ & (0.410, 0.437, 0.461) &  0.683 &  0.587 &  0.732 \\
D/SF & $ 4.0$ & $ 10^{3}$ & $ 1.6 \cdot 10^{3}$ &  128x128x256 & $ 5710 $ & (0.368, 0.400, 0.426) &  0.703 &  0.590 &  0.758 \\
D/SF & $ 4.0$ & $ 10^{3}$ & $ 3.2 \cdot 10^{3}$ &  256x256x256 & $ 3917 $ & (0.320, 0.357, 0.388) &  0.725 &  0.595 &  0.772 \\
L & $ 10^{-2}$ & $ 10^{3}$ & $ 8.0 \cdot 10^{2}$ &  128x128x256 & $ 1139 $ & (0.017, 0.030, 0.051) &  0.873 &  0.783 &  0.445 \\
L & $ 3.0 \cdot 10^{-2}$ & $ 10^{3}$ & $ 8.0 \cdot 10^{2}$ &  128x128x256 & $ 929  $ & (0.020, 0.044, 0.070) &  0.863 &  0.782 &  0.448 \\
L & $ 10^{-1}$ & $ 10^{3}$ & $ 8.0 \cdot 10^{2}$ &  128x128x256 & $ 1142 $ & (0.081, 0.076, 0.102) &  0.848 &  0.725 &  0.450 \\
L & $ 3.0 \cdot 10^{-1}$ & $ 10^{3}$ & $ 8.0 \cdot 10^{2}$ &  128x128x256 & $ 1109 $ & (0.076, 0.129, 0.157) &  0.825 &  0.655 &  0.451 \\
L & $1.0$ & $ 10^{3}$ & $ 8.0 \cdot 10^{2}$ &  128x128x256 & $ 3000 $ & (0.182, 0.225, 0.251) &  0.787 &  0.599 &  0.442 \\
L & $ 2.0$ & $ 10^{3}$ & $ 8.0 \cdot 10^{2}$ &  128x128x256 & $ 3000 $ & (0.278, 0.315, 0.340) &  0.759 &  0.570 &  0.436 \\
L & $ 4.0$ & $ 10^{3}$ & $ 8.0 \cdot 10^{2}$ &  128x128x256 & $ 10000$ & (0.399, 0.431, 0.455) &  0.737 &  0.518 &  0.428 \\
L & $ 8.0$ & $ 10^{3}$ & $ 8.0 \cdot 10^{2}$ &  128x128x256 & $ 5000 $ & (0.519, 0.545, 0.562) &  0.718 &  0.484 &  0.421 \\
L & $ 1.6 \cdot 10^{1}$ & $ 10^{3}$ & $ 8.0 \cdot 10^{2}$ &  128x128x256 & $ 8000 $ & (0.687, 0.709, 0.723) &  0.700 &  0.442 &  0.417 \\
\enddata                                                   	
\tablecomments{                          
Simulation type is specified as ``D'' for discontinuous/Case I or ``L'' for linear/Case II.
``D/SF'' simulations have stress-free boundary conditions.
Input control parameters are listed for each simulation: the penetration parameter $\mP$, stiffness $\mS$, and freefall Reynolds number $\mR$.
We also note the coefficient resolution (Chebyshev coefficients $nz$ and Fourier coeficients $nx$, $ny$).
We report the number of freefall time units each simulation was run for $t_{\rm{sim}}$.
Time-averaged values of the departure heights ($\delta_{0.1}$, $\delta_{0.5}$, $\delta_{0.9}$), the dissipation fraction $f$, and the dissipation fall-off $\xi$, as well as the average convection zone velocity $\angles{u}$ are reported.
We take these time averages over the final 1000 freefall times or half of the simulation, whichever is shorter.
}                                                          	
\end{deluxetable*}

\newpage
\,
\newpage
\bibliographystyle{aasjournal}
\bibliography{biblio}
\end{document}